\documentclass[pra,twocolumn,amsmath,amssymb,superscriptaddress]{revtex4-1}
\usepackage{epsfig,amsmath}
\usepackage{subfigure}
\usepackage{graphicx}% Include figure files
\usepackage{dcolumn}% Align table columns on decimal point
\usepackage{stmaryrd}
\usepackage{mathrsfs}
\usepackage{pifont}
\usepackage{amsthm}
\usepackage{amssymb}
\usepackage{bm}
\usepackage{latexsym}
\usepackage[colorlinks=true,linkcolor=blue,citecolor=blue]{hyperref}
\usepackage{color}
\usepackage{epstopdf}

\usepackage{booktabs}
\usepackage{threeparttable}
\usepackage{multirow}

\begin{document}

\title{Universal quantum control over non-Hermitian continuous-variable systems}

\author{Zhu-yao Jin}
\affiliation{School of Physics, Zhejiang University, Hangzhou 310027, Zhejiang, China}

\author{Jun Jing}
\email{Contact author: jingjun@zju.edu.cn}
\affiliation{School of Physics, Zhejiang University, Hangzhou 310027, Zhejiang, China}

\date{\today}

\begin{abstract}
Current studies about the continuous-variable systems in non-Hermitian quantum mechanics heavily revolved around the singularities in the eigenspectrum by mimicking their discrete-variable counterparts. Discussions over the nonunitary features in the time evolution are growing and yet limited in scalability and controllability. We develop here a general theory to control an arbitrary number of bosonic modes under the time-dependent non-Hermitian Hamiltonian. Far beyond the subspace of few excitations, our control theory operates in the Heisenberg picture and exploits the gauge potential underlying the instantaneous frames rather than the eigenspectrum. In particular, the instantaneous frames are defined by time-dependent ancillary operators as linear combinations of the laboratory-frame operators, while the associated gauge potential arises from the unitary transformation connecting the time-dependent and stationary ancillary frames. We find that the upper triangularization condition of the non-Hermitian Hamiltonian's coefficient matrix in the stationary ancillary frame gives rise to two nonadiabatic passages in both bra and ket spaces and also the exact solutions of the time-dependent Schr\"odinger equation. At the end of these passages, the probability conservation of the system wave function can be automatically restored without brute-force normalization. Our theory is exemplified by perfect and nonreciprocal state transfers in a cavity magnonic system under the non-Hermitian Hamiltonian rigorously derived from the Lindblad master equation with all quantum-jump terms retained. Under certain conditions, the perfect state transfer holds for arbitrary initial states and is irrelevant to both parity-time symmetry of the coefficient matrix and exceptional points of the eigenspectrum. The nonreciprocal transfer is consistent with the coherent perfect absorption. Our work promises a first-principles approach for the coherent control over non-Hermitian continuous-variable systems.
\end{abstract}

\maketitle

\section{Introduction}

Conventional control over non-Hermitian quantum systems is sensitive to spectrum, surrounding such as the broken or unbroken phases of parity-time ($\mathcal{PT}$) symmetry~\cite{Bender1998Real,Ashida2020NonHermitian,Bender2024PTsymmetric} of the Hamiltonian or scattering matrix~\cite{Metelmann2015Nonreciprocal,Ramezani2014Unidirectional,Stefano2015Unidirectional,
Jin2016IncidentDirection,Ramezani2016Unidirectional,Huang2017Unidirectional,Jin2018Incident,Harder2018Level,
Grigoryan2018Synchronized,Wang2019Nonreciprocity,Xu2024Robust} and the presence or absence of the exceptional points (EP)~\cite{Heiss2012Physics}. $\mathcal{PT}$ symmetry of the Hamiltonian provides a relaxed criterion for determining whether the eigenspectrum of the system is real or complex~\cite{Bender1998Real}. EPs describe the coalescence of eigenvalues and eigenstates~\cite{Heiss2012Physics}, where the biorthogonal condition breaks down~\cite{Brody2013Biorhogonal}. Spectrum-dependent protocols often impose stringent experimental requirements~\cite{Ibanez2011Shortcuts,Torosov2013NonHermitian,Torosov2014NonHermitian}. More seriously, the extraction of the eigenspectrum becomes increasingly challenging when the system extends from time-independent to time-dependent and from few dimensions to a large scale. The current work aims to establish a universal theory for non-Hermitian continuous-variable systems by a first-principles analysis rather than a phenomenological description of nonunitary dynamics~\cite{Metelmann2015Nonreciprocal,Ramezani2014Unidirectional,Stefano2015Unidirectional,
Jin2016IncidentDirection,Ramezani2016Unidirectional,Huang2017Unidirectional,Jin2018Incident,Harder2018Level,
Grigoryan2018Synchronized,Wang2019Nonreciprocity,Xu2024Robust}. It does not merely mimic its discrete-variable counterpart~\cite{Bender1998Real,Ashida2020NonHermitian,Bender2024PTsymmetric,Jin2025NonHerm} for a subspace of few excitations. Also it constitutes an indispensable chapter of our universal quantum control (UQC)~\cite{Jin2025Universal,Jin2025Entangling,Jin2025ErrCorr,Jin2025Rydberg,Jin2025Majorana}, beyond the closed continuous-variable systems~\cite{Jin2026Bosonic} and the Schr\"odinger picture.

Non-Hermitian Hamiltonian is a ubiquitous description for continuous-variable systems, such as photonic systems~\cite{Guo2009Observation,Regensburger2012Parity,Ge2013Antisymmetric,Peng2014Parity,Hodaei2017Enhanced,
Zhang2020Synthetic,Bergman2021Observation}, acoustic systems~\cite{Zhu2014PT,Fleury2015Invisible,Christensen2016Parity,Liu2018Unidirectional,Wang2019Extremely,Tang2020Exceptional}, and cavity magnonic systems~\cite{Zhang2017Observation,Zhang2019Experimental,Zhao2020Observation,Wang2022Dissipation,Qian2024Probing,
Yang2024Anomalous,Zhang2025Gain}. Typically it arises either from the unavoidable energy or material exchange with the environment~\cite{Metelmann2015Nonreciprocal,Ramezani2014Unidirectional,Stefano2015Unidirectional,Jin2016IncidentDirection,
Ramezani2016Unidirectional,Huang2017Unidirectional,Jin2018Incident,Harder2018Level,Grigoryan2018Synchronized,
Wang2019Nonreciprocity,Xu2024Robust} or from post-selection over no-quantum-jump trajectories~\cite{Han2024Measuring}, leading to the nonunitary evolution and the violation of probability conservation of the system wave function~\cite{Ashida2020NonHermitian}. Owing to the cooperative coupling to a common reservoir~\cite{Metelmann2015Nonreciprocal,Wang2019Nonreciprocity}, the non-Hermitian Hamiltonian accounts for the gain or loss effects associated with individual bosonic modes~\cite{Zhang2017Observation,Zhang2025Gain} and the dissipative coupling between the system modes~\cite{Metelmann2015Nonreciprocal,Harder2018Level,Grigoryan2018Synchronized,Wang2019Nonreciprocity}. In practice, the loss effect on a cavity mode can be effectively converted to the gain effect by applying external microwave fields to its ports~\cite{Zhang2017Observation,Zhang2025Gain}.

By mimicking the protocols developed for two-level systems~\cite{Ibanez2011Shortcuts,Torosov2013NonHermitian,Torosov2014NonHermitian,Chen2016Method,
Luan2022Shortcuts}, the shortcut-to-adiabaticity technique was directly applied to the non-Hermitian cavity magnonic systems~\cite{Zhang2022NonHermitian,Liu2025Nonadiabatic} in a two-dimensional subspace. Such spectrum-dependent protocols are typically related to time-dependent gain or loss rates of bosonic modes and need artificial renormalization of the system wave function. A scattering matrix based on the input-output theory~\cite{Zhang2017Observation,
Ramezani2014Unidirectional,Stefano2015Unidirectional,Metelmann2015Nonreciprocal,Jin2016IncidentDirection,
Ramezani2016Unidirectional,Huang2017Unidirectional,Jin2018Incident,Harder2018Level,Grigoryan2018Synchronized,
Wang2019Nonreciprocity,Xu2024Robust,Zou2024Dissipative} formulates a phenomenological and yet rough description for the non-Hermitian continuous-variable systems in the whole Hilbert space. Through precise match between coherent and engineered dissipative interaction, this approach can be used to demonstrate the directional quantum amplification~\cite{Metelmann2015Nonreciprocal}, the unidirectional microwave invisibility~\cite{Jin2018Incident,Wang2019Nonreciprocity}, and coherent perfect absorption~\cite{Ramezani2014Unidirectional,Stefano2015Unidirectional,Jin2016IncidentDirection,
Ramezani2016Unidirectional,Huang2017Unidirectional,Jin2018Incident,Xu2024Robust}. Most existing theories are constrained by the system size and the controllability of target states, excessively depending on the system spectrum, have to repeatedly renormalize the wave function during the time evolution and then cannot be regarded as a systematic and versatile framework for engineering the non-Hermitian bosonic systems.

The universal quantum control framework~\cite{Jin2025Universal,
Jin2025Entangling,Jin2025ErrCorr,Jin2025Rydberg,Jin2025NonHerm,Jin2025Majorana,Jin2026Bosonic} provides a much broader perspective by exploiting the gauge potential that emerges from the transformation to the time-independent (stationary) ancillary frames. In this paper, it is updated to copy with the multiple-mode bosonic systems governed by the non-Hermitian and time-dependent Hamiltonian that is rigorously obtained by the Lindblad master equation with the quantum-jump terms retained. We prove that the imposed triangularization condition of the coefficient matrix of a non-Hermitian Hamiltonian in the stationary frame enables exact solutions of the Heisenberg equation for two of the ancillary operators. They activate two nonadiabatic passages in both ket and bra spaces toward arbitrary target modes. In contrast to the conventional treatments about non-Hermitian systems~\cite{Daley2009Atomic,Uzdin2012Timedependent,
Ibanez2014Adiabaticity,Ashida2018Full-Counting,Dora2020Quantum}, our theory is objective, deterministic, and scalable in terms of control target and inherently conserves probability at the end of the passage without artificial normalization. In a cavity magnonic system with gain or loss effects on individual modes and a dissipative coupling between them, it is found that the triangularization condition always holds and then an arbitrary state in the cavity mode can be perfectly transferred to the magnon mode and unidirectional perfect absorption~\cite{Ramezani2014Unidirectional,Stefano2015Unidirectional,Jin2016IncidentDirection,Ramezani2016Unidirectional,
Huang2017Unidirectional,Jin2018Incident,Xu2024Robust} can be realized at the operational level if the initial states of these two modes are exchanged.

The rest of this paper is structured as follows. In Sec.~\ref{general}, we introduce a general theory for solving the Heisenberg equation for the bosonic operators driven by a non-Hermitian and time-dependent Hamiltonian. If the Hamiltonian's coefficient matrix can be triangularized in a stationary representation, then two ancillary operators can be decoupled from the others. In Sec.~\ref{Model}, the general theory is applied to the non-Hermitian cavity-magnonic system and, in Sec.~\ref{UQCpassage}, we construct the relevant Heisenberg passages, in which the state norm can be conserved by a proper parametric setting. Section~\ref{StateEP} demonstrates that an arbitrary state prepared in the cavity mode can be perfectly transferred to the magnon mode through a designed passage, which is irrelevant to the parity-time symmetry and the presence or absence of exceptional points. In addition, Sec.~\ref{CPA} shows that under the same setting a unidirectional perfect absorption occurs when the two modes are exchanged. The entire work is summarized in Sec.~\ref{conclusion}. Appendix~\ref{recipe} provides a brief recipe for constructing the ancillary operators as well as the unitary transformation between the time-dependent and stationary ancillary frames. Appendix~\ref{SuffTrian} provides the proof about the activation of two passages in the ket space from the upper-triangular Hamiltonian. Appendix~\ref{HeffDerive} details the derivation of the non-Hermitian Hamiltonian based on the master equation for the cavity-magnonic system. Appendix~\ref{Exam} illustrates the upper triangularization condition for the two-mode system.

\section{General framework}\label{general}

Consider a general bosonic system consisting of $N$ bosonic modes, associated with the annihilation operators $a_1$,$a_2$,$\ldots$,$a_N$, respectively. The system is controlled by a non-Hermitian Hamiltonian $H(t)$, i.e., $H(t)\ne H^\dagger(t)$. Under the assumption of the biorthogonal condition~\cite{Brody2013Biorhogonal}, namely that the bra and ket spaces are equipped with distinct bases, the system dynamics can be described by two sets of time-dependent Schr\"odinger equations as ($\hbar\equiv1$)
\begin{subequations}\label{Sch}
\begin{align}
i\frac{d}{dt}|\psi(t)\rangle&=H(t)|\psi(t)\rangle,\label{SchH}\\
i\frac{d}{dt}|\phi(t)\rangle&=H^\dagger(t)|\phi(t)\rangle, \label{SchHd}
\end{align}
\end{subequations}
where $|\psi(t)\rangle$ and $\langle\phi(t)|$ are the pure-state solutions in the ket and bra spaces, respectively. For the quadratic bosonic systems, the time-dependent Hamiltonian $H(t)$ and its Hermitian conjugate $H^\dagger(t)$ can be expressed as
\begin{equation}\label{Ham}
H(t)=\vec{a}^\dagger H^a(t)\vec{a}^T, \quad H^\dagger(t)=\vec{a}^\dagger [H^a(t)]^\dagger\vec{a}^T,
\end{equation}
where $\vec{a}\equiv (a_1,a_2,\ldots,a_N)$ and $\vec{a}^\dagger=(a_1^\dagger,a_2^\dagger,\ldots,a_N^\dagger)$ are row vectors of operator and $H^a(t)$ is an $N\times N$ time-dependent coefficient matrix with Hermitian conjugate $[H^a(t)]^\dagger$. $H^a(t)\neq[H^a(t)]^\dagger$. The superscript $T$ denotes the matrix transposition and transforms a row vector to its corresponding column vector. Due to the noncommunity, non-Hermicity, and the infinite-dimensional Hilbert space of the time-dependent Hamiltonian, solving the Schr\"odinger equation for non-Hermitian continuous-variable systems is generally challenging. However, our UQC theory~\cite{Jin2025Universal,Jin2025Entangling,Jin2025ErrCorr,
Jin2025Rydberg,Jin2025NonHerm,Jin2025Majorana,Jin2026Bosonic} can be extended to the non-Hermitian bosonic systems, providing a fundamental framework to partially solve Eqs.~(\ref{SchH}) and (\ref{SchHd}) in the same time.

Without loss of generality, we start with the system dynamics in the ket space, as given by Eq.~(\ref{SchH}) with $H(t)$ in Eq.~(\ref{Ham}). Inspired by the universal quantum control for Hermitian bosonic systems~\cite{Jin2026Bosonic}, the system dynamics can be described in the ancillary representation associated with a completed set of time-dependent ancillary operators $\mu_k(t)$, $1\leq k\leq N$. $\{\mu_k(t)\}$ are superposed of the laboratory-frame bosonic operators $\{a_k\}$ with an $N\times N$ unitary transformation matrix $\mathcal{M}^\dagger(t)$:
\begin{equation}\label{TimeAnci}
\vec{\mu}_t^T=\mathcal{M}^\dagger(t)\vec{a}^T, \quad \vec{\mu}_t\equiv[\mu_1(t), \mu_2(t), \ldots, \mu_N(t)].
\end{equation}
A general $\mathcal{M}^\dagger(t)$ is constructed in appendix~\ref{recipe}. Using Eq.~(\ref{TimeAnci}), one can confirm that the ancillary operators $\mu_k(t)$ satisfy the canonical communication relation, i.e., $[\mu_j(t), \mu_k^\dagger(t)]=\delta_{jk}$. Consequently, the time-dependent Hamiltonian~(\ref{Ham}) can be formulated as
\begin{equation}\label{HamTimeAnci}
H(t)=\vec{\mu}_t^\dagger H^\mu(t)\vec{\mu}_t^T,
\end{equation}
where
\begin{equation}\label{HamCoeffMu}
H^\mu(t)=\mathcal{M}^\dagger(t)H^a(t)\mathcal{M}(t)
\end{equation}
is the Hamiltonian's coefficient matrix in the representation of time-dependent ancillary operators.

To proceed, we consider the rotation from the time-dependent ancillary operators to the time-independent or stationary ancillary operators, i.e., $\vec{\mu}_t\rightarrow\vec{\mu}_0$ with $\vec{\mu}_0=[\mu_1(0), \mu_2(0), \ldots, \mu_N(0)]$. Such a rotation can be achieved by the unitary transformation $\mathcal{V}(t)$ as $\mathcal{V}^\dagger(t)\mu_k(t)\mathcal{V}(t)=\mu_k(0)$, where $\mathcal{V}(t)$ is determined by  $\mathcal{M}^\dagger(t)$ in Eq.~(\ref{TimeAnci}) (see appendix~\ref{recipe} for details). Generally, in the rotating frame with respect to $\mathcal{V}(t)$, we have
\begin{equation}\label{HamTimeInAnci}
\begin{aligned}
H_{\rm rot}(t)&=\mathcal{V}^\dagger(t)H(t)\mathcal{V}(t)-i\mathcal{V}^\dagger(t)\frac{d\mathcal{V}(t)}{dt}\\
&=\vec{\mu}_0^\dagger\left[H^\mu(t)-\mathcal{A}(t)\right]\vec{\mu}_0^T=\vec{\mu}_0^\dagger\mathcal{H}(t)\vec{\mu}_0^T,
\end{aligned}
\end{equation}
where the coefficient matrix in the stationary ancillary representation $\mathcal{H}(t)\equiv H^\mu(t)-\mathcal{A}(t)$, with the non-Hermitian dynamical coefficient matrix $H^\mu(t)\neq[H^\mu(t)]^\dagger$. The Hermitian and purely geometric matrix $\mathcal{A}$ represents the gauge potential~\cite{Michael2017Geometry} associated with the unitary transformation $\mathcal{V}(t)$. The element of the matrix $\mathcal{A}$ in the $k$th row and $m$th column is defined as $\mathcal{A}_{km}=-i[\mu_k^\dagger(t), d\mu_m(t)/dt]$. In parallel, for the bra-space dynamics in Eq.~(\ref{SchHd}), the Hamiltonian expressed by the stationary ancillary operators is the Hermitian conjugate of $H_{\rm rot}(t)$, i.e., $H^\dagger_{\rm rot}(t)=\vec{\mu}_0^\dagger[(H^\mu(t))^\dagger-\mathcal{A}(t)]\vec{\mu}_0^T=\vec{\mu}_0^\dagger\mathcal{H}^\dagger(t)\vec{\mu}_0^T$.

Consequently, the time-dependent Schr\"odinger equations~(\ref{SchH}) and (\ref{SchHd}) are transformed as
\begin{subequations}\label{Schrot}
\begin{align}
i\frac{d}{dt}|\psi(t)\rangle_{\rm rot}&=H_{\rm rot}(t)|\psi(t)\rangle_{\rm rot},\label{SchHrot}\\
i\frac{d}{dt}|\phi(t)\rangle_{\rm rot}&=H^\dagger_{\rm rot}(t)|\phi(t)\rangle_{\rm rot}, \label{SchrotHd}
\end{align}
\end{subequations}
respectively, with the rotated pure states
\begin{equation}\label{Purerot}
|\psi(t)\rangle_{\rm rot}=\mathcal{V}^\dagger(t)|\psi(t)\rangle,\quad |\phi(t)\rangle_{\rm rot}=\mathcal{V}^\dagger(t)|\phi(t)\rangle.
\end{equation}
The evolution operators for $|\psi(t)\rangle_{\rm rot}$ and $|\phi(t)\rangle_{\rm rot}$ are
\begin{equation}\label{Usch}
U_{\rm rot}(t)=\hat{T}e^{-i\int_0^tH_{\rm rot}(s)ds},\quad V_{\rm rot}(t)=\hat{T}e^{-i\int_0^tH_{\rm rot}^\dagger(s)ds},
\end{equation}
respectively, where $\hat{T}$ is the time-ordering operator.

The rotation to the stationary representation~(\ref{HamTimeInAnci}) does not directly relieve the difficulty in solving the non-Hermitian Schr\"odinger equation~(\ref{Schrot}). On one hand, it is not appropriate to apply the commutation condition about the Hamiltonian's coefficient matrix and the projection operator with time-independent ancillary modes established for the Hermitian bosonic network~\cite{Jin2026Bosonic}, since the non-Hermitian coefficient matrix $\mathcal{H}(t)$ is generally nondiagonalizable, i.e., $\mathcal{H}_{km}(t)\ne\mathcal{H}_{mk}(t)$ for $k\ne m$. On the other hand, the triangularization condition about the rotated Hamiltonian for discrete-variable systems~\cite{Jin2025NonHerm} also fails here due to the fundamentally distinct statistics of bosonic modes.

\emph{Main result.} We prove that the upper triangularization of the coefficient matrix of $\mathcal{H}(t)$ is a sufficient condition for activating two useful Heisenberg-picture passages for non-Hermitian continuous-variable systems (see Appendix~\ref{SuffTrian} for detailed derivations). They emerge from the differential manifold and seem not relevant to the spectral characteristics. Specifically, the upper triangularization condition for the coefficient matrix $\mathcal{H}(t)$ can be briefly expressed by~\cite{Jin2025NonHerm,Ju2024Emergent}
\begin{equation}\label{UpTriaEq}
\mathcal{H}(t)\Pi^k-\Pi^k\mathcal{H}^T(t)=0,
\end{equation}
where $\Pi^k$, $1\leq k\leq N$, is the projection operator or an $N\times N$ matrix defined by $\Pi^k_{jm}=\delta_{jk}\delta_{mk}$. Alternatively, the upper triangularization condition (\ref{UpTriaEq}) means the upper triangularized Hamiltonian:
\begin{equation}\label{HamrotDiaUp}
\begin{aligned}
H_{\rm rot}(t)&=\sum_{k=1}^N\sum_{m\geq k}^N\left[H^\mu_{km}(t)-\mathcal{A}_{km}(t)\right]\mu_k^\dagger(0)\mu_m(0)\\
&=\sum_{k=1}^N\sum_{m\geq k}^N\mathcal{H}_{km}(t)\mu_k^\dagger(0)\mu_m(0).
\end{aligned}
\end{equation}
In Appendix~\ref{SuffTrian}, we demonstrate that the Hamiltonian~(\ref{HamrotDiaUp}) constitutes a sufficient condition to decouple the ancillary operators $\mu_1^\dagger(0)$ and $\mu_N(0)$ from the others. Their dynamics in the original picture can be written as
\begin{equation}\label{DycAncHN}
\mu_1^\dagger(0)\rightarrow e^{if_1(t)}\mu_1^\dagger(t), \quad \mu_N(0)\rightarrow e^{-if_N(t)}\mu_N(t),
\end{equation}
where the global phases are defined as
\begin{equation}\label{globalf}
f_k(t)=\int_0^t\mathcal{H}_{kk}(s)ds, \quad k=1,N.
\end{equation}
Along the two passages indicated by $\mu_1^\dagger(t)$ and $\mu_N(t)$, the systems initially prepared in the states $F[\mu_1^\dagger(0)]|{\rm vac}\rangle$ and $F[\mu_N(0)]|{\rm vac}\rangle$ will evolve to the desired target modes $F[e^{if_1(t)}\mu_1^\dagger(t)]|{\rm vac}\rangle$ and $F[e^{-if_N(t)}\mu_N(t)]|{\rm vac}\rangle$ at the time $t$ with the accumulated global phases $f_1(t)$ and $-f_N(t)$, respectively, where $F[\cdot]$ represents an arbitrary function of operators.

In parallel, for the systems governed by the Hermitian-conjugate Hamiltonian $H_{\rm rot}^\dagger(t)$ in the bra (dual) space, the upper triangularization condition~(\ref{UpTriaEq}) can activate $\mu_1(t)$ and $\mu_N^\dagger(t)$ as nonadiabatic passages in the Heisenberg picture. In this case, the Hermitian conjugate of Eq.~(\ref{HamrotDiaUp}) takes the lower triangularization form of
\begin{equation}\label{HamrotDiaLower}
H^\dagger_{\rm rot}(t)=\sum_{k=1}^N\sum_{m\geq k}^N\mathcal{H}_{km}^*(t)\mu_m^\dagger(0)\mu_k(0).
\end{equation}
Similar to the ket-space case, the dynamics of the ancillary operators $\mu_1(0)$ and $\mu_N^\dagger(0)$ can be obtained as
\begin{equation}\label{DycAncHd}
\mu_1(0)\rightarrow e^{-if_1^*(t)}\mu_1(t), \quad \mu_N^\dagger(0)\rightarrow e^{if_N^*(t)}\mu_N^\dagger(t)
\end{equation}
by using the non-Hermitian Heisenberg equation and the Hamiltonian~(\ref{HamrotDiaLower}), where the global phase $f_k^*(t)$, $k=1,N$, is the complex conjugate of $f_k(t)$ in Eq.~(\ref{globalf}).

\section{Non-Hermitian cavity-magnonic system}\label{NHcavity}

\subsection{Model and Hamiltonian}\label{Model}

\begin{figure}[htbp]
\centering
\includegraphics[width=0.8\linewidth]{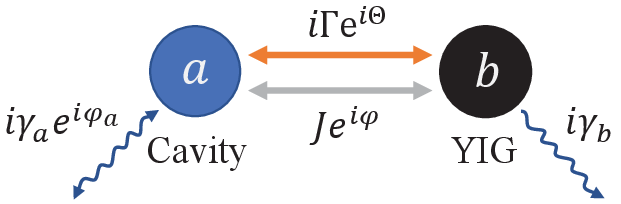}
\caption{Sketch of an open cavity magnonic system under control. More than the coherent exchange coupling with strength $J$ and phase $\varphi$ between the cavity mode $a$ and the magnon mode $b$, the cooperative coupling of the two-mode system to the environment gives rise to the gain or loss rate $\gamma_a$ of the cavity mode, the loss rate $\gamma_b$ of the magnon mode, and the dissipation coupling $i\Gamma e^{i\Theta}$ between them. $\Theta$ can be $0$ or $\pi$ in experiments~\cite{Wang2019Nonreciprocity}.}\label{model}
\end{figure}

In this section, we apply our UQC theory for the general non-Hermitian bosonic systems in Sec.~\ref{general} to analyze an open two-mode bosonic system that consists of a single-mode cavity and an yttrium iron garnet sphere in the Kittel mode~\cite{Harder2018Level,Grigoryan2018Synchronized,Bhoi2019Abnormal,Yang2019Control,Yao2019Microscopic,Yu2019Prediction,
Boventer2020Control} as shown in Fig.~\ref{model}. The cavity and magnon modes are represented by the annihilation operators $a$ and $b$, respectively. They are mutually interacted by the time-dependent exchange coupling $J(t)$ with phase $\varphi$. The cooperative coupling between the two-mode system and the environment~\cite{Metelmann2015Nonreciprocal,Wang2019Nonreciprocity} leads to the gain or loss rates $\gamma_a$ and $\gamma_b$ of the cavity mode $a$ and the magnon mode $b$, respectively, as well as a dissipative coupling $i\Gamma e^{i\Theta}$ between them. Then the whole system can be described by a non-Hermitian Hamiltonian~\cite{Metelmann2015Nonreciprocal,Wang2019Nonreciprocity}:
\begin{equation}\label{HamMC1}
\begin{aligned}
&H(t)=\left(\omega_a-i\gamma_ae^{i\varphi_a}\right)a^\dagger a+\left[\omega_b(t)-i\gamma_b\right]b^\dagger b\\
+&\left[J(t)e^{i\varphi}+i\Gamma e^{i\Theta}\right]a^\dagger b
+\left[J(t)e^{-i\varphi}+i\Gamma e^{i\Theta}\right]b^\dagger a.
\end{aligned}
\end{equation}
Here $\omega_a$ is the free frequency of the cavity mode. $\omega_b(t)$ is the time-dependent frequency of the magnon modes, which can be flexibly tuned by the external bias magnetic field $B(t)$~\cite{Zhang2014Strongly,Babak2022Cavity,Jin2024Stabilizing}. $\gamma_b>0$ is a constant loss rate of magnon mode. $\gamma_a>0$ represents the gain or loss effect on the cavity mode when $\varphi_a=\pi$ or $\varphi_a=0$, respectively~\cite{Zhang2017Observation,Zhang2025Gain}. The phase $\Theta=0$ or $\Theta=\pi$ of the dissipative coupling is determined by the propagation direction of the traveling waves~\cite{Wang2019Nonreciprocity}. For simplicity, we set $\Theta=0$ in the following. The derivation details about the non-Hermitian Hamiltonian~(\ref{HamMC1}) through the adjoint Lindblad master equation, i.e., the Lindblad master equation in the Heisenberg picture, can be found in Appendix~\ref{HeffDerive} with $\varphi_a=0$, despite both $\varphi_a$ and $\Gamma$ having many choices~\cite{Wang2019Nonreciprocity,Zhang2017Observation,Zhang2025Gain}. We capture the full non-Hermitian Hamiltonian without omitting the quantum-jump terms. Thus the Schr\"odinger equation with the time-dependent non-Hermitian Hamiltonian~(\ref{HamMC1}) is equivalent to the Lindblad master equation.

In the rotating frame with respect to $H_0(t)=\omega_0(t)(a^\dagger a+b^\dagger b)$, the Hamiltonian~(\ref{HamMC1}) is transformed as
\begin{equation}\label{HamMC}
\begin{aligned}
H(t)&=\left[\frac{\Delta(t)}{2}-i\gamma_ae^{i\varphi_a}\right]a^\dagger a-\left[\frac{\Delta(t)}{2}+i\gamma_b\right]b^\dagger b\\
&+\left[J(t)e^{i\varphi}+i\Gamma\right]a^\dagger b+\left[J(t)e^{-i\varphi}+i\Gamma\right]b^\dagger a,
\end{aligned}
\end{equation}
where the detuning $\Delta(t)$ follows $\Delta(t)/2=\omega_a-\omega_0(t)=-\omega_b(t)+\omega_0(t)$. One can find that the system Hamiltonian~(\ref{HamMC}) satisfies $\mathcal{PT}$ symmetry~\cite{Ashida2020NonHermitian,Bender2024PTsymmetric} under the setting of $\varphi_a=\pi$, $\gamma_a=\gamma_b$, $\Delta(t)=0$, and the vanishing dissipative coupling $\Gamma=0$. In fact, the Hamiltonian eigenenergies
\begin{equation}\label{eigenenergy}
\begin{aligned}
&E_\pm=-i\frac{\gamma_ae^{i\varphi_a}+\gamma_b}{2}\pm\frac{1}{2}
\Big\{\Delta^2-\left(\gamma_ae^{i\varphi_a}-\gamma_b\right)^2\\ &+4(J^2-\Gamma^2)
+i\left[8J\Gamma\cos\varphi-2\Delta\left(\gamma_ae^{i\varphi_a}-\gamma_b\right)\right]\Big\}^{1/2}.
\end{aligned}
\end{equation}
become real-valued, i.e., $E_\pm=\pm\sqrt{J^2-\gamma_a^2}$, when $[H(t), \mathcal{PT}]=0$ and $J>\gamma_a$. Otherwise, they are complex and the Hamiltonian is in the $\mathcal{PT}$-symmetry-broken phase. Whether holding $\mathcal{PT}$ symmetry or not, the system can approach EP under the conditions of $\Delta(t)=0$, $\varphi=\pi/2$, $\gamma_a=\gamma_b$, and $J=\pm\Gamma$ (when $\varphi_a=0$) or $J=\pm\sqrt{\Gamma^2+\gamma_a^2}$ (when $\varphi_a=\pi$).

Conventionally, the controls over non-Hermitian systems are based on their spectrum characteristics~\cite{Ashida2020NonHermitian,Bender2024PTsymmetric}, obsessing about $\mathcal{PT}$ symmetry and EPs. In contrast, our UQC theory in Sec.~\ref{general} is essentially proposed in the instantaneous frames that are irrespective to the spectral properties. In the current example, we will demonstrate that an arbitrary state can be perfectly transferred from the cavity mode to the magnon mode, regardless of whether the system is in the unbroken or broken phases of $\mathcal{PT}$ symmetry and whether EPs are present or not.

\subsection{Universal Heisenberg passages of two operators}\label{UQCpassage}

With Eq.~(\ref{HamMC}), we construct the universal passages for the non-Hermitian cavity-magnonic system, which give rise to nonunitary evolution of the system state and the final state can be automatically normalized. According to Eqs.~(\ref{unitary}) and (\ref{unitarVec}), the time-dependent ancillary operators for a two-mode system can be constructed as
\begin{equation}\label{AnciTwo}
[\mu_1(t), \mu_2(t)]^T=\mathcal{M}^\dagger(t)(a, b)^T
\end{equation}
with a $2\times2$ unitary transformation matrix
\begin{equation}\label{Mtwo}
\mathcal{M}^\dagger(t)=\begin{pmatrix}\cos\theta(t)
e^{i\frac{\alpha(t)}{2}}&-\sin\theta(t)e^{-i\frac{\alpha(t)}{2}}\\ \sin\theta(t)e^{i\frac{\alpha(t)}{2}}&\cos\theta(t)e^{-i\frac{\alpha(t)}{2}}\end{pmatrix},
\end{equation}
where the parameters $\theta(t)$ and $\alpha(t)$ manipulate the population and the relative phase of both cavity mode $a$ and magnon mode $b$, respectively. By Eqs.~(\ref{unitaryV}) and (\ref{uniVat}), the unitary transformation $\mathcal{V}(t)$ in Eq.~(\ref{HamTimeInAnci}), which acts as $\mathcal{V}^\dagger(t)\mu_1(t)\mathcal{V}(t)=\mu_1(0)$ and $\mathcal{V}^\dagger(t)\mu_2(t)\mathcal{V}(t)=\mu_2(0)$, can take the form of
\begin{equation}\label{unitaryVtwo}
\mathcal{V}(t)=V_\alpha(t)V_\theta(t),
\end{equation}
with
\begin{equation}\label{uniVattwo}
\begin{aligned}
V_\alpha(t)&=e^{-i\frac{\alpha(t)-\alpha(0)}{2}(a^\dagger a-b^\dagger b)},\\
V_\theta(t)&=e^{-[\theta(t)-\theta(0)]\left[e^{i\alpha(0)}b^\dagger a-e^{-i\alpha(0)}a^\dagger b\right]}.
\end{aligned}
\end{equation}
It is interesting to find that the preceding transformation from the time-dependent ancillary operators to the time-independent ones can also be formally implemented by $[V_\theta(t)V_\alpha(t)]^\dagger\mu_k(t)V_\theta(t)V_\alpha(t)\rightarrow\mu_k(0)$, $k=1,2$, despite the fact that $[V_\alpha(t), V_\theta(t)]\neq0$.

Substituting the Hamiltonian~(\ref{HamMC}) and Eqs.~(\ref{AnciTwo}--\ref{uniVattwo}) into the upper triangularization condition~(\ref{UpTriaEq}), we obtain the constraints for the coupling strength $J(t)$ and the detuning $\Delta(t)$ as
\begin{equation}\label{Hconstr}
\begin{aligned}
J(t)&=\Big[\dot{\theta}(t)+\Gamma\cos\alpha(t)\cos2\theta(t)-\left(\gamma_a\cos\varphi_a-\gamma_b\right)\\
&\times\sin\theta(t)\cos\theta(t)\Big]/\sin[\varphi+\alpha(t)],\\
\Delta(t)&=\dot{\alpha}(t)-2\Big[J(t)\cos(\varphi+\alpha(t))\cot2\theta(t)\\
&+\Gamma\frac{\sin\alpha(t)}{\sin2\theta(t)}+\frac{\gamma_a\sin\varphi_a}{2}\Big].
\end{aligned}
\end{equation}
The details can be found in Appendix~\ref{Exam}. At the operational level, the singularities of $J(t)$ and $\Delta(t)$ should be avoided by appropriately choosing the parameters $\varphi$, $\alpha(t)$, and $\theta(t)$. Under Eq.~(\ref{Hconstr}), the ancillary operator $\mu_1^\dagger(t)$ can be activated as a nonadiabatic Heisenberg passage that does not necessarily conserve the state norm. According to Eq.~(\ref{DycAncHN}), we have
\begin{equation}\label{AnciTwoAct}
\mu_1^\dagger(0)\rightarrow e^{if_1(t)}\mu_1^\dagger(t),
\end{equation}
where the complex global phase $f_1(t)$ can be divided into the real part $f_r(t)$ and the imaginary part $f_i(t)$ as
\begin{equation}\label{globaltwo}
\dot{f}_1(t)=\dot{f}_r(t)+\dot{f}_i(t)
\end{equation}
with
\begin{equation}\label{globaltwoRI}
\begin{aligned}
\dot{f}_r(t)&=\frac{1}{2}\Delta(t)\cos2\theta(t)-J(t)\cos\left[\varphi+\alpha(t)\right]
\sin2\theta(t)\\ &-\frac{\dot{\alpha}(t)}{2}\cos2\theta(t)+\gamma_a\sin\varphi_a\cos^2\theta(t),\\
\dot{f}_i(t)&=-i\Big[\gamma_a\cos\varphi_a\cos^2\theta(t)+\gamma_b\sin^2\theta(t) \\
&-\Gamma\cos\alpha\sin2\theta(t)\Big].
\end{aligned}
\end{equation}
Equations~(\ref{AnciTwo}), (\ref{Mtwo}), and (\ref{AnciTwoAct}) indicate that the time evolution of the two-bosonic-mode system is subject to the boundary conditions of $\theta(t)$ and $\alpha(t)$. For example, when $\theta(0)=0$ and $\theta(\tau)=\pi/2$ with $\tau$ the evolution period, $\mu_1^\dagger(0)=a^\dagger\rightarrow\mu_1^\dagger(\tau)=b^\dagger$, and then the initial state of the cavity mode $a$ can be perfectly transferred along the passage $\mu_1^\dagger(t)$ to the magnon mode $b$ that is prepared as the vacuum state. In addition, the imaginary part of the phase $\dot{f}_i(t)$ in Eq.~(\ref{globaltwoRI}) captures the fact that the non-Hermitian component in the Hamiltonian~(\ref{HamMC}) renders the probability nonconservation during the time evolution. However, in our protocol, the state norm can be guaranteed to be unit at both beginning and end of the evolution, by imposing the vanishing integral $\int_0^\tau \dot{f}_i(t)dt=f_i(\tau)-f_i(0)=0$.

Despite the fact that this condition does not directly emerge from the dynamics, it is a mild restriction and can be always attainable if we have a sufficient number of tunable parameters. For example, with constant $\gamma_a$, $\gamma_b$, $\varphi_a$, $\alpha$, and $\dot{\theta}(t)$, the time integral of $\dot{f}_i(t)$ in Eq.~(\ref{globaltwoRI}) is
\begin{equation}\label{imgfint}
\begin{aligned}
f_i(\tau)-f_i(0)&=-\frac{i}{2}(\gamma_a\cos\varphi_a+\gamma_b)\tau-\frac{i}{2\dot{\theta}(t)}\\
&\times\Big\{(\gamma_a\cos\varphi_a-\gamma_b)[\sin2\theta(\tau)-\sin2\theta(0)]\\
&+\Gamma\cos\alpha[\cos2\theta(\tau)-\cos2\theta(0)]\Big\},
\end{aligned}
\end{equation}
which always vanishes under the condition of
\begin{equation}\label{vanif}
\begin{aligned}
\Gamma=&-\Big[(\gamma_a\cos\varphi_a+\gamma_b)\tau\dot{\theta}(t)+(\gamma_a\cos\varphi_a-\gamma_b)
\Big(\sin2\theta(\tau)\\
&-\sin2\theta(0)\Big)\Big]/\Big[\cos\alpha\Big(\cos2\theta(\tau)-\cos2\theta(0)\Big)\Big].
\end{aligned}
\end{equation}
Note $\Gamma$ is a constant number determined by the linear function $\theta(t)$.

In parallel, for the system dynamics governed by the Hermitian conjugate $H^\dagger(t)$ in the dual space, the same conditions in Eq.~(\ref{Hconstr}) can activate the ancillary operator $\mu_2^\dagger(t)$ as the nonadiabatic passage. In particular, the time evolution takes the form of Eq.~(\ref{DycAncHd}), where the global phase $\dot{f}_2^*(t)=i[\gamma_a\exp(-i\varphi_a)+\gamma_b]-\dot{f}_1^*(t)$ with $\dot{f}_1^*(t)$ the complex conjugate of $\dot{f}_1(t)$ given by Eqs.~(\ref{globaltwo}) and (\ref{globaltwoRI}). Similar to $\mu_1^\dagger(t)$, a flexible and perfect state transfer can be implemented along the passage $\mu_2^\dagger(t)$ under appropriate choices of $\theta(t)$ and $\alpha(t)$. The probability conservation can also be ensured in the end of the evolution.

\section{State control over cavity and magnon}\label{Statetransfer}

\subsection{Perfect state transfer}\label{StateEP}

In this section, we use the activated passage $\mu_1^\dagger(t)$ in Eq.~(\ref{AnciTwoAct}) to realize the perfect transfer of arbitrary initial states from the cavity mode to the magnon mode, including the Fock state, the binomial code state (a state of a logical qubit encoded for enhancing noise resilience)~\cite{Michael2016NewClass}, the coherent state, the cat state, and even the thermal state. In our system, these transfers are found to be irrelevant to the $\mathcal{PT}$ symmetry of the Hamiltonian and the presence or absence of EPs.

We first consider the conditions of $\varphi_a=\pi$, $\gamma_a=\gamma_b$, and $\Gamma=0$, corresponding to the balanced gain and loss effects in cavity and magnon modes with vanishing dissipative coupling. Under these conditions, the constraints in Eq.~(\ref{Hconstr}) together with the global phase in Eq.~(\ref{globaltwoRI}) reduce to
\begin{equation}\label{HconstrRed}
\begin{aligned}
J(t)=&\left[\dot{\theta}(t)+\gamma_a\sin2\theta(t)\right]/\sin[\varphi+\alpha(t)],\\
\Delta(t)=&\dot{\alpha}(t)-2J(t)\cos[\varphi+\alpha(t)]\cot2\theta(t),\\
\dot{f}_r(t)=&\frac{1}{2}\Delta(t)\cos2\theta(t)-J(t)\cos\left[\varphi+\alpha(t)\right]\sin2\theta(t)\\
&-\frac{\dot{\alpha}(t)}{2}\cos2\theta(t),\\
\dot{f}_i(t)=&i\gamma_a\cos2\theta(t).
\end{aligned}
\end{equation}
If we further set $\dot{\alpha}(t)=0$ and $\varphi+\alpha(t)=\pi/2$, then $\Delta(t)=0$ and the Hamiltonian~(\ref{HamMC}) is in the unbroken phase of $\mathcal{PT}$ symmetry.

Using Eqs.~(\ref{AnciTwo}), (\ref{Mtwo}), and (\ref{AnciTwoAct}), the system passage can evolve as $\mu_1^\dagger(0)=a^\dagger\rightarrow e^{if_r(\tau)}\mu_1^\dagger(\tau)=e^{if_r(\tau)}b^\dagger$ under the setting of $\alpha(t)=0$,
\begin{equation}\label{paratheta}
\theta(t)=\frac{\pi t}{2\tau},
\end{equation}
and the imposed condition $f_i(\tau)-f_i(0)=0$. The last condition ensures that the system state probability is conserved at the end of the evolution, regardless of the presence or absence of exceptional points. To have a vanishing time integral over $\dot{f}_i(t)$ in Eq.~(\ref{HconstrRed}), one can choose the gain or loss rate as
\begin{equation}\label{condvaniGam}
\gamma_a=\lambda\dot{\theta}(t),
\end{equation}
given $\theta(t)$ as the linear function of time in Eq.~(\ref{paratheta}). Here the positive factor $\lambda$ scales the rate magnitude.

Assume that the cavity mode and magnon mode are initially prepared in the Fock state $|n=5\rangle$ and the vacuum state $|0\rangle$, respectively, i.e.,
\begin{equation}
|\psi(0)\rangle=|5\rangle_a|0\rangle_b=\frac{(a^\dagger)^5}{\sqrt{5!}}|0\rangle_a|0\rangle_b
=\frac{[\mu_1^\dagger(0)]^5}{\sqrt{5!}}|0\rangle_a|0\rangle_b.
\end{equation}
It will be finally transformed to be
\begin{equation}\label{initrans}
\frac{[\mu_1^\dagger(\tau)]^5}{\sqrt{5!}}|0\rangle_a|0\rangle_b
=\frac{(b^\dagger)^5}{\sqrt{5!}}|0\rangle_a|0\rangle_b=|0\rangle_a|5\rangle_b=|\psi(\tau)\rangle
\end{equation}
up to an irrelevant phase, demonstrating a perfect Fock-state transfer from the cavity mode to the magnon mode.

The performance of our protocol for state transfer can be evaluated by the fidelity $\mathcal{F}_{\rho}=\langle\psi(t)|\rho|\psi(t)\rangle$, where $|\psi(t)\rangle$ is the pure-state solution of the time-dependent Schr\"odinger equation~(\ref{SchH}) with the Hamiltonian~(\ref{HamMC}). $\rho$ is the density matrix of any interested state, such as initial, intermediate, and target states. For pure states $\rho=|\phi\rangle\langle\phi|$, we have $\mathcal{F}_{\phi}=|\langle\phi|\psi(t)\rangle|^2$. When the initial state is a product of Fock states, it is equivalent to show the population dynamics $\mathcal{F}_{n_1,n_2}=|\langle n_1|\langle n_2|\psi(t)\rangle|^2$ during the time evolution.

\begin{figure}[htbp]
\centering
\includegraphics[width=0.9\linewidth]{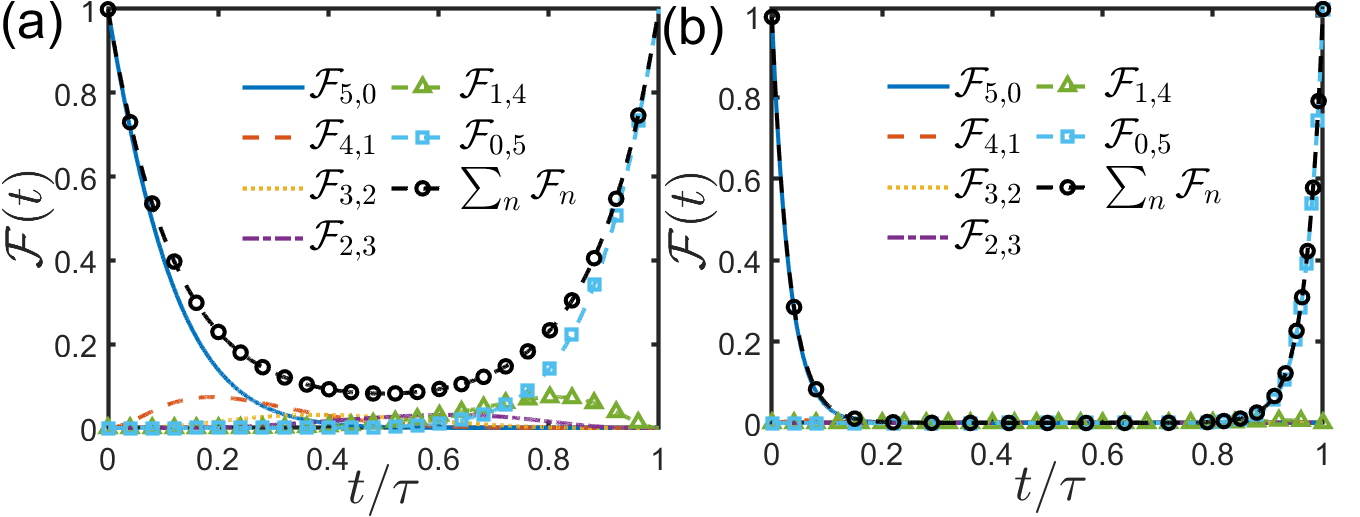}
\includegraphics[width=0.9\linewidth]{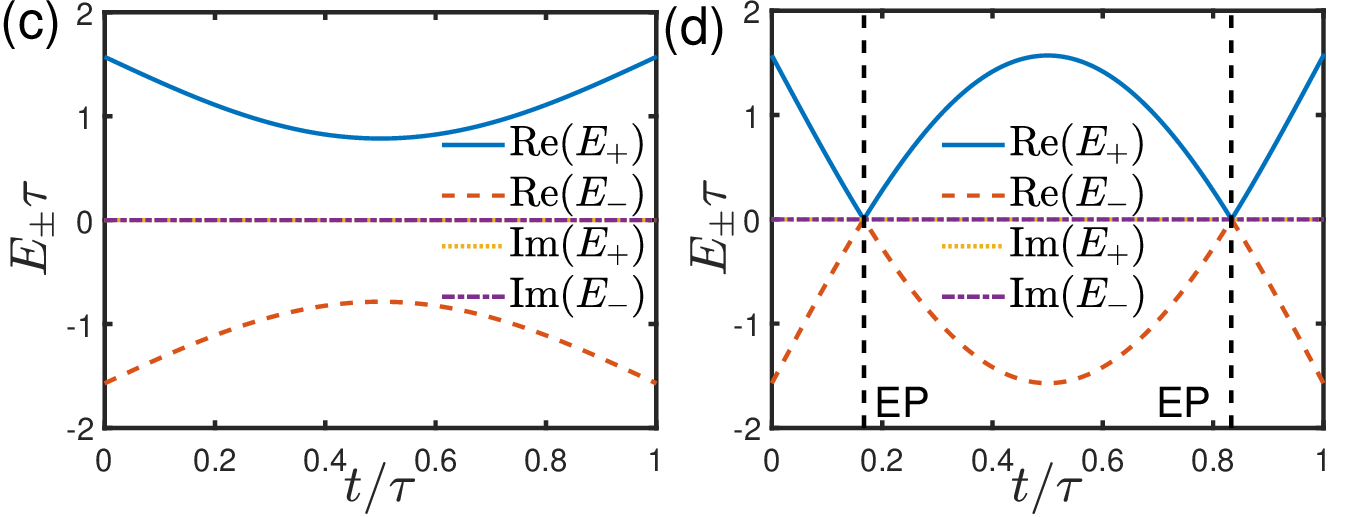}
\caption{Fidelity dynamics during the Fock-state transfer $|\psi(0)\rangle=|5\rangle_a|0\rangle_b\rightarrow|\psi(\tau)\rangle=|0\rangle_a|5\rangle_b$ under the $\mathcal{PT}$-symmetric Hamiltonian, using the passage $\mu_1^\dagger(t)$ in the cavity-magnonic system for (a) avoiding EPs and (b) crossing EPs. The associated dynamics of the real and imaginary parts of the energies $E_\pm$ in Eq.~(\ref{eigenenergy}) is plotted in (c) and (d) for avoiding and crossing EPs, respectively. With $\theta(t)$ in Eq.~(\ref{paratheta}), the coherent coupling strength $J(t)$ and the detuning $\Delta(t)$ are constrained by Eq.~(\ref{HconstrRed}) where $\gamma_a$ satisfies Eq.~(\ref{condvaniGam}) with $\lambda=1$ in (a) and (c) and $\lambda=4$ in (b) and (d). Under this parametric setting, we have constants $\gamma_a=\gamma_b=\pi/2\tau$ in (a) and (c) and $\gamma_a=\gamma_b=2\pi/\tau$ in (b) and (d). And $f_i(\tau)-f_i(0)=0$ for both avoiding and crossing EPs.}\label{ConverFockPT}
\end{figure}

Figures~\ref{ConverFockPT}(a) and (b) demonstrate perfect passage $\mu_1^\dagger(t)$, i.e., $\mathcal{F}_{5,0}(0)=1\rightarrow\mathcal{F}_{0,5}(\tau)=1$, under a $\mathcal{PT}$-symmetric Hamiltonian. It is found that for both evolutions of (a) avoiding and (b) crossing EPs, the Fock states of conserved excitations other than the initial and target states are rarely populated. For example, in Fig.~\ref{ConverFockPT}(a), we have at most $\mathcal{F}_{4,1}(0.19\tau)=0.073$, $\mathcal{F}_{3,2}(0.40\tau)=0.031$, $\mathcal{F}_{2,3}(0.62\tau)=0.032$, and $\mathcal{F}_{1,4}(0.82\tau)=0.074$. The total fidelity $\sum_n\mathcal{F}_n\equiv\sum_{n_1,n_2}\mathcal{F}_{n_1,n_2}$, being equivalent to the trace of the two-mode system ${\rm Tr}[|\psi(t)\rangle\langle\psi(t)|]$, can represent the probability conservation or nonconservation under a non-Hermitian Hamiltonian. In both Figs.~\ref{ConverFockPT}(a) and (b), it is found that $\sum_n\mathcal{F}_n(0<t<\tau)<1$ due to the gain effect of the cavity mode and the loss effect of the magnon mode. Nevertheless, $\sum_n\mathcal{F}_n(t=\tau)$ always converges to unit at the end of the passage. In Figs.~\ref{ConverFockPT}(c) and (d), we present the dynamics of real and imaginary parts of eigenenergies $E_\pm$ due to Eq.~(\ref{eigenenergy}), relevant to the avoiding and crossing EPs in the parametric setting, respectively. The curves in Fig.~\ref{ConverFockPT}(c) verify that the system dynamics avoids the coalescence of eigenenergies and eigenstates. Figure~\ref{ConverFockPT}(d) indicates that the system crosses EPs when $t=0.18\tau$ and $t=0.82\tau$. For the two-mode system discussed in this work, it is hard to find a clear connection between the perfect state transfer and the existence of the exceptional points of the system spectrum.

\begin{figure}[htbp]
\centering
\includegraphics[width=0.9\linewidth]{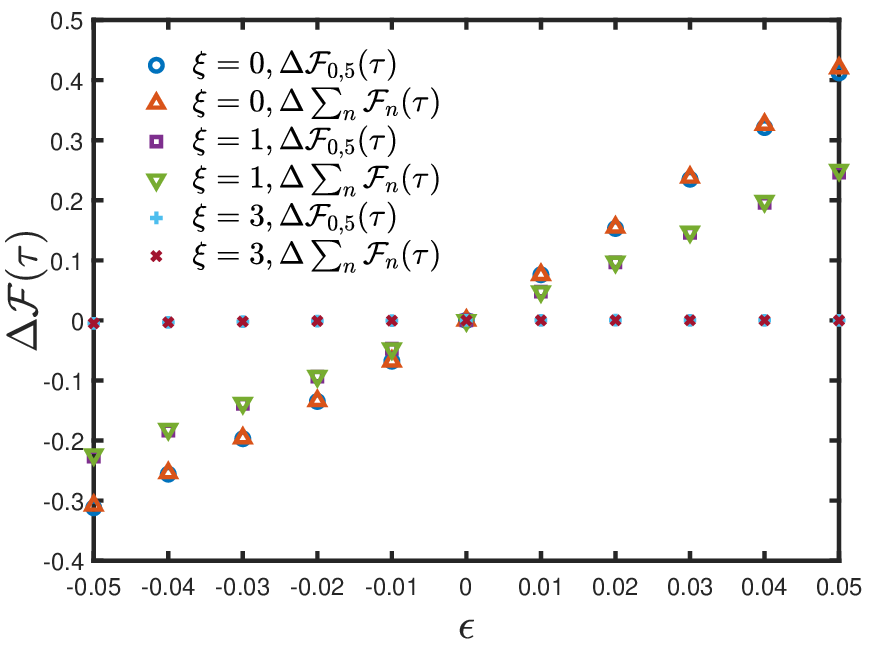}
\caption{Fidelity difference $\Delta\mathcal{F}_{0,5}(\tau)$ for the target state $|0\rangle_a|5\rangle_b$ and total fidelity difference $\Delta\sum_n\mathcal{F}_n(\tau)$ between the ideal and nonideal situations as functions of the deviation coefficient $\epsilon$ under various control coefficient $\xi$, during the Fock-state transfer $|\psi(0)\rangle=|5\rangle_a|0\rangle_b\rightarrow|\psi(\tau)\rangle=|0\rangle_a|5\rangle_b$. $J(t)$ and $\Delta(t)$ are set as Eq.~(\ref{HconstrThetaf}). $f_r(t)$ is set by Eq.~(\ref{setf}). The other parameters are the same as Fig.~\ref{ConverFockPT}(a).}\label{deltaF}
\end{figure}

Our protocol is sensitive to the parametric deviation in the absence of error correction. Without loss of generality, we consider that the control Hamiltonian~(\ref{HamMC}) is under the fluctuation of the coherent interaction between the two modes, i.e.,
\begin{equation}\label{HamErr}
H(t)\rightarrow H(t)+\epsilon H_1(t), \quad H_1(t)=J(t)e^{i\varphi}a^\dagger b+{\rm H.c.},
\end{equation}
where $\epsilon$ scales the deviation magnitude. With $\epsilon\neq0$, the constraints in Eq.~(\ref{Hconstr}) or Eq.~(\ref{HconstrRed}) found by the ideal Hamiltonian and the relevant upper triangularization condition are no longer valid. Subsequently, the imposed condition $f_i(\tau)-f_i(0)=0$ will fail to ensure the probability conservation of the system at the desired moment $t=\tau$. It can be confirmed by the blue circles and the yellow upper triangles in Fig.~\ref{deltaF} with $\xi=0$, which demonstrate the target-state fidelity difference $\Delta\mathcal{F}_{0,5}(\tau)\equiv\mathcal{F}_{0,5}'(\tau)-\mathcal{F}_{0,5}(\tau)$ and the total fidelity difference $\Delta\sum_n\mathcal{F}_n(\tau)\equiv\sum_n[\mathcal{F}_n'(\tau)-\mathcal{F}_n(\tau)]$ with the nonideal fidelity $\mathcal{F}_n'(\tau)$, respectively. Specifically, the target-state fidelity difference and total population difference are as large as $\Delta\mathcal{F}_{0,5}(\tau)=-0.314$ and $\Delta\sum_n\mathcal{F}_n(\tau)=-0.308$ for $\epsilon=-0.05$, $\Delta\mathcal{F}_{0,5}(\tau)\sim\Delta\sum_n\mathcal{F}_n(\tau)=-0.068$ for $\epsilon=-0.01$, $\Delta\mathcal{F}_{0,5}(\tau)\sim\Delta\sum_n\mathcal{F}_n(\tau)=0.076$ for $\epsilon=0.01$, and $\Delta\mathcal{F}_{0,5}(\tau)=0.420$ and $\Delta\sum_n\mathcal{F}_n(\tau)=0.412$ for $\epsilon=0.05$.

According to the dynamical-correction strategy~\cite{Jin2025ErrCorr} under the UQC framework, the unwanted leakage rate of the desired passage induced by the systematic errors scales as $\sim\int_0^\tau\exp{\Big(if_r(t)\Big)}P[\theta(t)]dt$, where $P[\theta(t)]$ is a complicated function of $\theta(t)$. When $P[\theta(t)]$ is a slowly varying function with time in comparison to the exponential function $\exp{\Big(if_r(t)\Big)}$, the unwanted leakage can be significantly suppressed. This can be simply realized by the parametric setting~\cite{Jin2025ErrCorr}
\begin{equation}\label{setf}
\dot{f}_r(t)=\xi\dot{\theta}(t),
\end{equation}
with a control coefficient $\xi>1$. At the operational level, the correction mechanism in Eq.~(\ref{setf}) can be implemented by rewriting $J(t)$ and $\Delta(t)$ in Eq.~(\ref{HconstrRed}) as functions of $\theta(t)$ and $f_r(t)$:
\begin{equation}\label{HconstrThetaf}
\begin{aligned}
J(t)&=\left[\dot{\theta}(t)+\gamma_a\sin2\theta(t)\right]\sqrt{1+\cot^2(\varphi+\alpha(t))},\\
\Delta(t)&=\dot{\alpha}(t)+2\dot{f}_r(t)\cos2\theta(t),\\
\dot{\alpha}(t)&=-\frac{\ddot{\tilde{\theta}}\dot{f}_r\sin2\theta-\dot{\tilde{\theta}}\ddot{f}_r\sin2\theta
-2\dot{\tilde{\theta}}^2\dot{f}_r\cos2\theta}{\dot{\tilde{\theta}}^2+\dot{f}_r^2\sin^22\theta}
\end{aligned}
\end{equation}
with $\dot{\tilde{\theta}}(t)\equiv\dot{\theta}(t)+\gamma_a\sin2\theta(t)$. The purple squares, the green down triangles, the light-blue pluses, and the dark-purple crosses in Fig.~\ref{deltaF} illustrate the effect of our correction mechanism against the systematic errors. In particular, for $\xi=3$, we have $|\Delta\mathcal{F}_{0,5}(\tau)|\sim|\Delta\sum_n\mathcal{F}_n(\tau)|\le0.003$ in the whole regime of $\epsilon\in[-0.05,0.05]$.

\begin{figure}[htbp]
\centering
\includegraphics[width=0.9\linewidth]{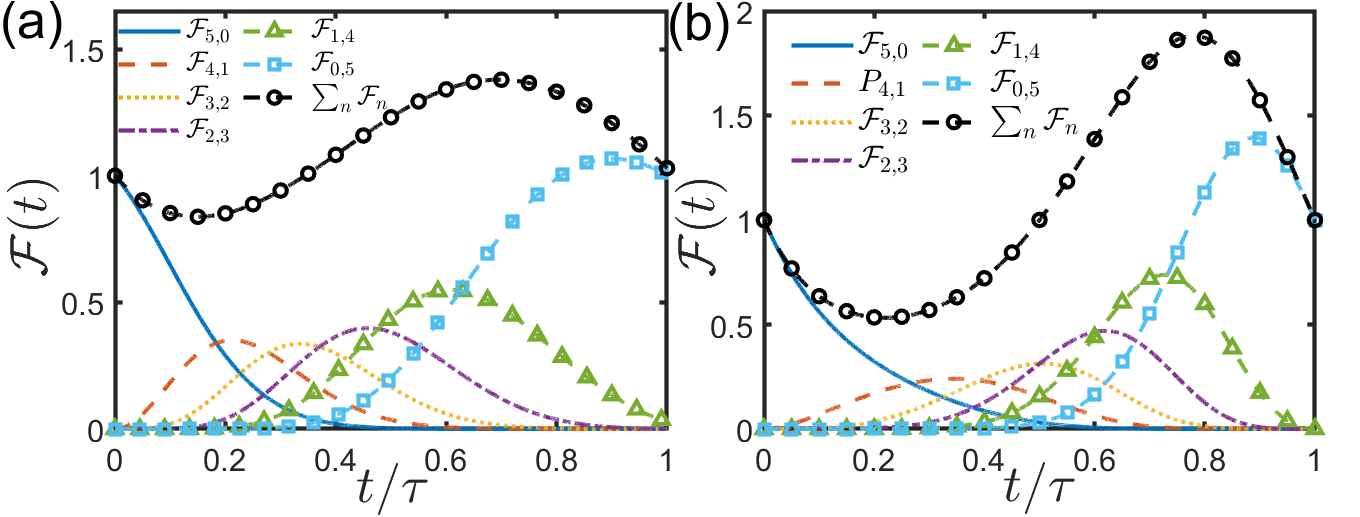}
\includegraphics[width=0.9\linewidth]{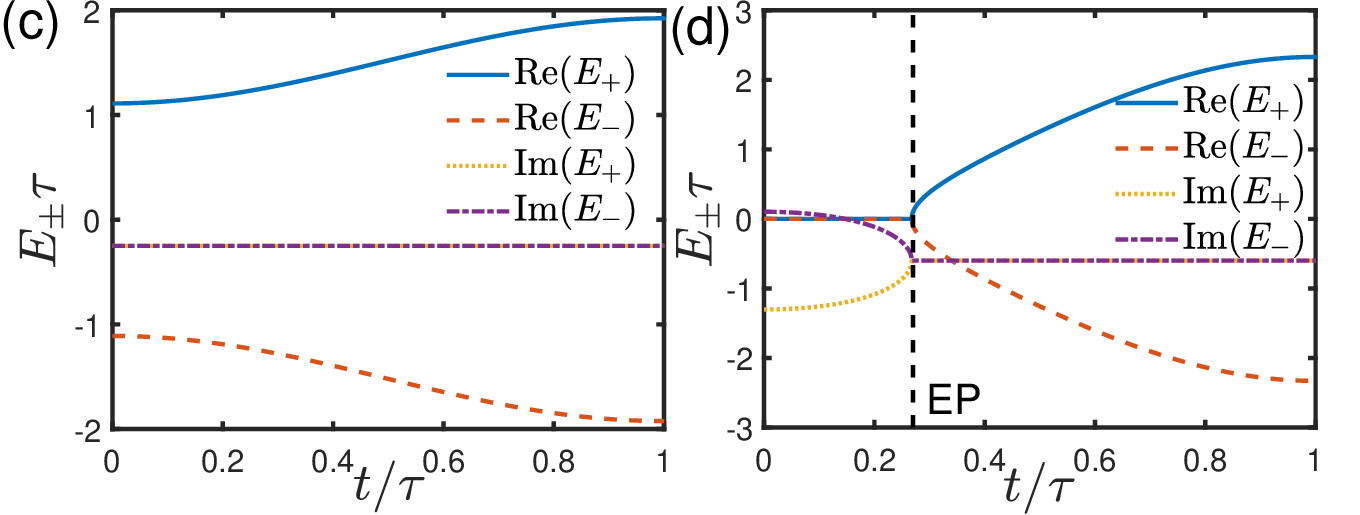}
\caption{Fidelity dynamics during the Fock-state transfer $|\psi(0)\rangle=|5\rangle_a|0\rangle_b\rightarrow|\psi(\tau)\rangle=|0\rangle_a|5\rangle_b$ under a $\mathcal{PT}$-symmetric-broken Hamiltonian, using the passage $\mu_1^\dagger(t)$ in the cavity-magnonic system for (a) avoiding EPs and (b) crossing EPs. The associated dynamics of the real and imaginary parts of the energies $E_\pm$ in Eq.~(\ref{eigenenergy}) is shown in (c) and (d) for avoiding and crossing EPs, respectively. With $\theta(t)$ in Eq.~(\ref{paratheta}), the coherent coupling strength $J(t)$ and the detuning $\Delta(t)$ are constrained by Eq.~(\ref{HconstrRed}) where $\varphi_a=0$ and $\gamma_a=\gamma_b$, with $\gamma_a$ and $\Gamma$ set by Eq.~(\ref{condvanifi}), using $\lambda=0.5$ in (a) and (c) and $\lambda=1.2$ in (b) and (d). $f_i(\tau)-f_i(0)=0$ for both avoiding and crossing EPs. }\label{ConverFock}
\end{figure}

Similar to Figs.~\ref{ConverFockPT}(a) and (b), Figs.~\ref{ConverFock}(a) and (b) demonstrate the fidelity dynamics $\mathcal{F}_{n_1,n_2}$ of various five-excitation product states and their summation $\sum_n\mathcal{F}_n=\sum_{n_1,n_2}\mathcal{F}_{n_1,n_2}$ within the broken phase of $\mathcal{PT}$ symmetry in the absence and presence of EPs, respectively. Here we set $\varphi_a=0$, $\gamma_a=\gamma_b$, and $\Gamma\neq0$, relevant to loss effects in both the cavity and magnon modes with nonvanishing dissipative coupling. Under these conditions, the Hamiltonian is in the broken phase of $\mathcal{PT}$ symmetry. Then the imaginary part of the global phase in Eq.~(\ref{globaltwoRI}) becomes
\begin{equation}\label{globaltwoIred}
\dot{f}_i(t)=-i(\gamma_a-\Gamma\sin2\theta).
\end{equation}
To neutralize the imaginary phase, one can choose the loss rate and dissipative coupling strength as
\begin{equation}\label{condvanifi}
\gamma_a=\frac{\lambda}{\pi}\dot{\theta}(t),\quad \Gamma=-\frac{\lambda}{2}\dot{\theta}(t),
\end{equation}
with $\theta(t)$ in Eq.~(\ref{paratheta}). One can verify that $\gamma_a$ and $\Gamma$ are constants under the setting in Eq.~(\ref{condvanifi}).

In both Figs.~\ref{ConverFock}(a) and (b), it is found that the initial population on $|5\rangle_a|0\rangle_0$ can be completely transferred to $|0\rangle_a|5\rangle_b$, even when the other five-excitation states could be temporally yet significantly populated during the passage. The Fock states with more excitations of the target (magnon) mode are subsequently populated until $\mathcal{F}_{0,5}(\tau)=1$ at the end of the time evolution. For $0<t<\tau$, the summation of individual populations is not conserved, i.e., $\sum_n\mathcal{F}_n<1$ or $\sum_n\mathcal{F}_n>1$, due to the loss effects of both the cavity and magnon modes and the dissipative coupling between them. The peak value of the population summation is found to be $\sum_n\mathcal{F}_n=1.32$ when $t=0.72\tau$ in Fig.~\ref{ConverFock}(a), around which the dissipative coupling between the cavity and magnon modes dominates the loss of both modes. In Fig.~\ref{ConverFock}(a) with no EPs, we have $\mathcal{F}_{4,1}(0.22\tau)=0.35$, $\mathcal{F}_{3,2}(0.34\tau)=0.34$, $\mathcal{F}_{2,3}(0.45\tau)=0.40$, $\mathcal{F}_{1,4}(0.61\tau)=0.55$, and $\mathcal{F}_{0,5}(0.90\tau)=1.07$. And, in Fig.~\ref{ConverFock}(b) with EPs, we have $\mathcal{F}_{4,1}(0.35\tau)=0.24$, $\mathcal{F}_{3,2}(0.50\tau)=0.32$, $\mathcal{F}_{2,3}(0.61\tau)=0.46$, $\mathcal{F}_{1,4}(0.72\tau)=0.74$, and $\mathcal{F}_{0,5}(0.89\tau)=1.40$. Under Eq.~(\ref{condvanifi}), the state-norm becomes unit at the end of the passage. Again the associated evolutions of the real and imaginary parts of the eigenenergies are presented in Figs.~\ref{ConverFock}(c) and (d), respectively. As shown in Fig.~\ref{ConverFock}(c), the eigenenergies do not coalesce during the whole passage, confirming the avoidance of EPs. In Fig.~\ref{ConverFock}(d), an EP occurs at $t=0.27\tau$, which is also irrelevant to the system dynamics.

\begin{figure}[htbp]
\centering
\includegraphics[width=0.9\linewidth]{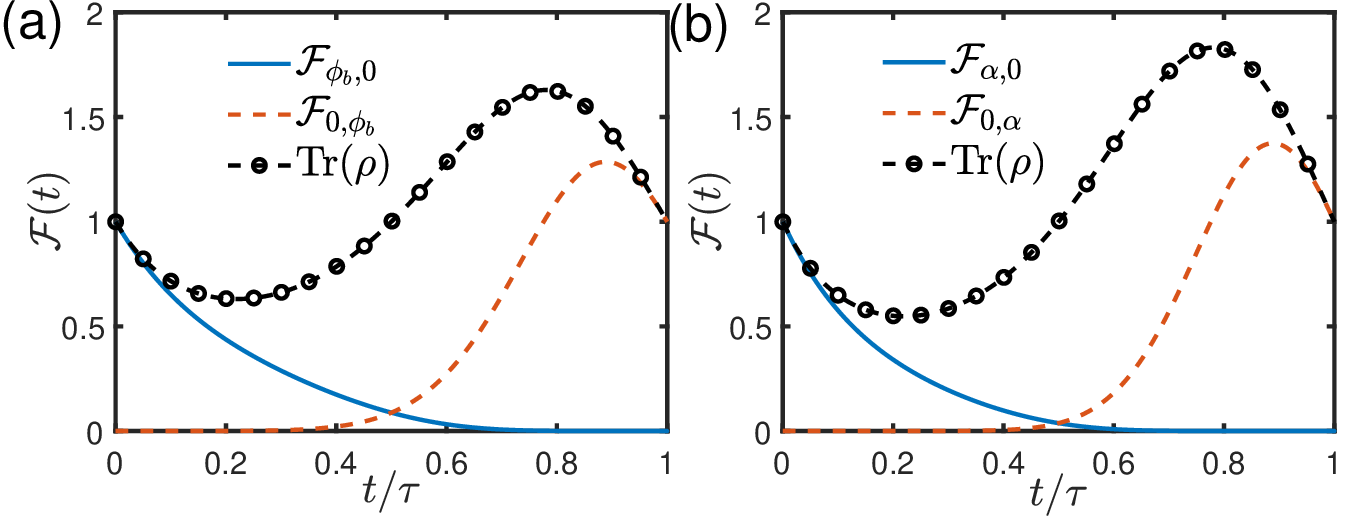}
\includegraphics[width=0.9\linewidth]{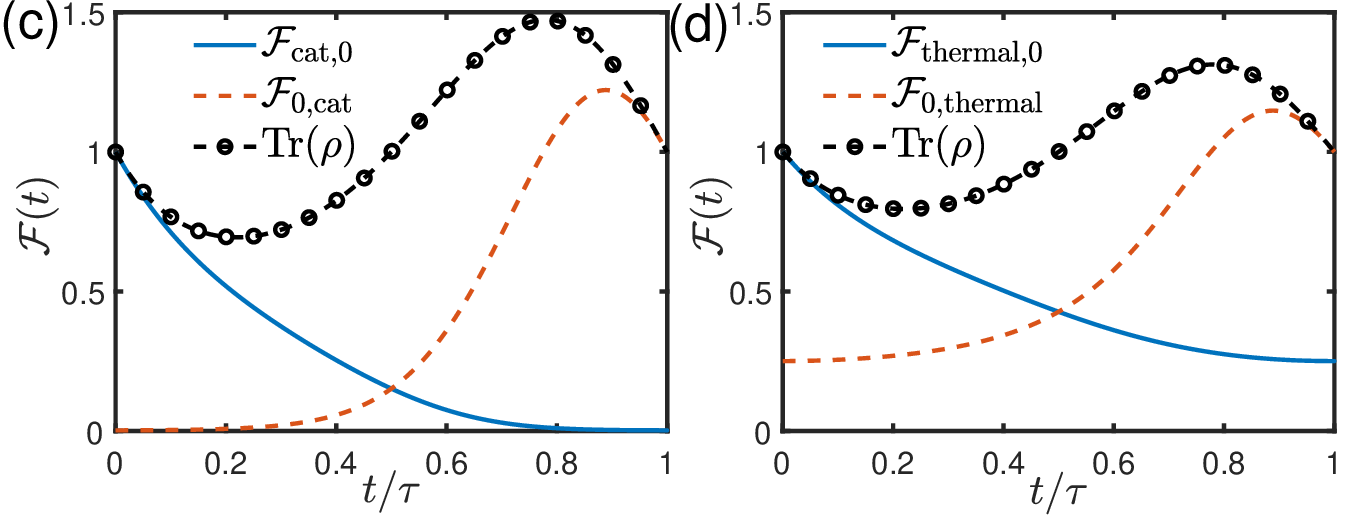}
\caption{Fidelity dynamics during the perfect transfer of (a) the binomial code state $|\phi_b,0\rangle\rightarrow|0,\phi_b\rangle$ with $|\phi_b\rangle=(\sqrt{3}|2\rangle+|6\rangle)/2$~\cite{Michael2016NewClass}, (b) the coherent state $|\alpha,0\rangle\rightarrow|0,\alpha\rangle$ with $\alpha=5$, (c) the cat state $|\rm cat,0\rangle\rightarrow|0,\rm cat\rangle$, where $|\rm cat\rangle=(|\alpha\rangle+|-\alpha\rangle)/\sqrt{2}$ with $\alpha=5$, and (d) the thermal state $\rho_{\rm th}\otimes|0\rangle\langle0|\rightarrow|0\rangle\langle0|\otimes\rho_{\rm th}$, where $\rho_{\rm th}=\sum_np_n|n\rangle\langle n|$ with $p_n=(\bar{n}^n)/(1+\bar{n})^{n+1}$ and $\bar{n}=5$. The parameters are the same as Figs.~\ref{ConverFock}(b) and (d).}\label{biono}
\end{figure}

Far beyond the Fock state, the universal Heisenberg passage can unravel perfect transfer of diverse states. In Fig.~\ref{biono}(a), the cavity and magnon modes are initially prepared in the binomial code state $|\phi_b\rangle=(\sqrt{3}|2\rangle+|6\rangle)/2$~\cite{Michael2016NewClass} and the vacuum state $|0\rangle$, respectively. The performance of our protocol can be evaluated by $\mathcal{F}_{\phi_b,0}=|\langle\phi_b|\langle0|\psi(t)\rangle|^2$ and $\mathcal{F}_{0,\phi_b}=|\langle0|\langle\phi_b|\psi(t)\rangle|^2$. It is found that the binomial code state in the cavity mode can be perfectly transferred to the magnon mode with a unit fidelity. Along the passage, the non-Hermitian system during $0<t<\tau$ is not subject to the probability conservation. For example, we have ${\rm Tr}(\rho)=0.631$ when $t=0.234\tau$ and ${\rm Tr}(\rho)=1.63$ when $t=0.782$. However, the probability conservation is restored as ${\rm Tr}(\rho)=1$ when $t=\tau$. Similarly, for the cavity modes initially prepared as the coherent state $|\alpha\rangle$ with $\alpha=5$, the cat state $|{\rm cat}\rangle=(|\alpha\rangle+|-\alpha\rangle)/\sqrt{2}$ with $\alpha=5$, and the thermal state $\rho_{\rm th}$ with $\bar{n}=5$ in Figs.~\ref{biono}(b), (c), and (d), respectively, all of them can be perfectly transferred to the magon mode. Note for the thermal state, $\mathcal{F}_{{\rm thermal},0}$ and $\mathcal{F}_{0,{\rm thermal}}$ are not complementary to each other in any situation.

A quantum state initially prepared in the magnon mode can be perfectly transferred to the cavity mode by activating the passage $\mu_2^\dagger(t)$ in the dual space. The parameter setting is almost the same as that for the passage $\mu_1^\dagger(t)$, except that $\varphi=\pi/2$ is replaced with $\varphi=-\pi/2$.

\subsection{Unidirectional perfect absorber}\label{CPA}

In this section, we show that the cavity-magnonic system under our UQC theory in Sec.~\ref{StateEP} can simulate a nonreciprocal behavior under the same control conditions as in Fig.~\ref{ConverFock} for $\mu_1^\dagger(t)$ in the ket space. For example, when the Fock state is initially prepared in the magnon mode, i.e., $|\psi(0)\rangle=|0\rangle_a|5\rangle_b$, the total population of the system is found to be nearly vanishing at the end of time evolution. To a certain degree, this unidirectional absorption can be interpreted by the coupled-mode theory~\cite{Xia2025Observation} about the coherent perfect absorber~\cite{Zhang2017Observation,Chong2010Coherent,
Wan2011TimeReversed,Baranov2017Coherent,Chen2020Perfect,Seunghwi2025Complex,Xia2025Observation}.

Under the conditions of Eq.~(\ref{Hconstr}) for $\Delta(t)$ with $\varphi_a=0$, $\varphi=\pi/2$, $\alpha(t)=0$, and $\gamma_a=\gamma_b$, the non-Hermitian Hamiltonian in Eq.~(\ref{HamMC}) turns out to be
\begin{equation}\label{HamMCRe}
\begin{aligned}
H(t)&=\begin{pmatrix}a^\dagger&b^\dagger\end{pmatrix}H^a(t)\begin{pmatrix}a\\b\end{pmatrix}\\
&=\begin{pmatrix}a^\dagger&b^\dagger\end{pmatrix}\begin{pmatrix}-i\gamma_a&iJ(t)+i\Gamma\\
-iJ(t)+i\Gamma&-i\gamma_a\end{pmatrix}\begin{pmatrix}a\\b\end{pmatrix}.
\end{aligned}
\end{equation}
In the coupled-mode theory~\cite{Xia2025Observation}, one has to consider the dissipation of the cavity-magnonic system to its external channels~\cite{Zhang2017Observation,Chong2010Coherent,
Wan2011TimeReversed,Baranov2017Coherent,Chen2020Perfect,Seunghwi2025Complex,Xia2025Observation}. This dissipation induces the external loss of the cavity and magnon modes~\cite{Zhang2017Observation}, by which the total loss rates of the cavity and magnon modes in Eq.~(\ref{HamMCRe}) can be divided as $\gamma=\gamma_0+\gamma_1$, where $\gamma_0$ and $\gamma_1$ represent the intrinsic and external loss rates of the modes, respectively. In our system, $\gamma=\gamma_a$ or $\gamma_b$.

Following the coupled-mode theory~\cite{Xia2025Observation}, the system dynamics can be described by the scattering matrix~\cite{Zhang2017Observation,Chong2010Coherent,Wan2011TimeReversed,Baranov2017Coherent,Chen2020Perfect,
Seunghwi2025Complex,Xia2025Observation}. It is defined as
\begin{equation}\label{Scatter}
S(\omega,t)=I-iK^\dagger\frac{1}{\omega-H^a(t)}K,
\end{equation}
where $K=\sqrt{2\gamma_1}I$ and $I$ is the two-dimensional identity operator. Under the assumption that the incident monochromatic acoustic wave~\cite{Xia2025Observation} is resonant with the cavity-magnonic system, i.e., $\omega=0$, the scattering matrix $S$ for $H(t)$ in Eq.~(\ref{HamMCRe}) can be written as
\begin{equation}\label{SpeScatter}
S(t)=\begin{pmatrix}S_{11}(t)&S_{12}(t)\\S_{21}(t)&S_{22}(t)\end{pmatrix},
\end{equation}
with the reflection coefficients $S_{11}(t)$ and $S_{22}(t)$ and the transmission coefficients $S_{12}(t)$ and $S_{21}(t)$ given by
\begin{equation}\label{matrixEle}
\begin{aligned}
S_{11}(t)&=S_{22}(t)=1+\frac{2\gamma_1\gamma_a}{D(t)},\\
S_{12}(t)&=\frac{2\gamma_1[J(t)+\Gamma]}{D(t)},\quad S_{21}(t)=-\frac{2\gamma_1[J(t)-\Gamma]}{D(t)},
\end{aligned}
\end{equation}
where $D(t)\equiv-\gamma_a^2-[J^2(t)-\Gamma^2]$.

With the same parameters as in Fig.~\ref{ConverFock}, i.e., $J(t)$, $\theta(t)$, and $\gamma_a$ and $\Gamma$ given by Eqs.~(\ref{Hconstr}), (\ref{paratheta}), and (\ref{condvanifi}), respectively, one can find that, at the end of the running period, $J(\tau)=\dot{\theta}-\Gamma\approx-\Gamma$ when $\lambda\approx1$. Consequently, the scattering matrix with  $t=\tau$ in Eq.~(\ref{Scatter}) becomes
\begin{equation}\label{SpeScatterT}
S(t=\tau)\approx\begin{pmatrix}1-\frac{2\gamma_1}{\gamma_a}&0\\S_{21}(\tau)&1-\frac{2\gamma_1}{\gamma_a}\end{pmatrix}
=\begin{pmatrix}0&0\\S_{21}(\tau)&0\end{pmatrix},
\end{equation}
where $S_{21}(\tau)\approx4\gamma_1J(\tau)/\gamma_a^2\neq0$. Here the second equivalence holds when $\gamma_1=\gamma_a/2$. Equation~(\ref{SpeScatterT}) indicates that for a nonvacuum initial state in the magnon mode $b$, the total system population can become almost zero at a desired moment, i.e., $[a(\tau), b(\tau)]^T=S(t=\tau)[0, b(0)]^T\approx(0, 0)^T$. In practice, the non-Hermitian cavity magnonic system now becomes a promising candidate~\cite{Ramezani2014Unidirectional,Stefano2015Unidirectional,Jin2016IncidentDirection,
Ramezani2016Unidirectional,Huang2017Unidirectional,Jin2018Incident,Xu2024Robust} for the unidirectional perfect absorber.

\begin{figure}[htbp]
\centering
\includegraphics[width=0.9\linewidth]{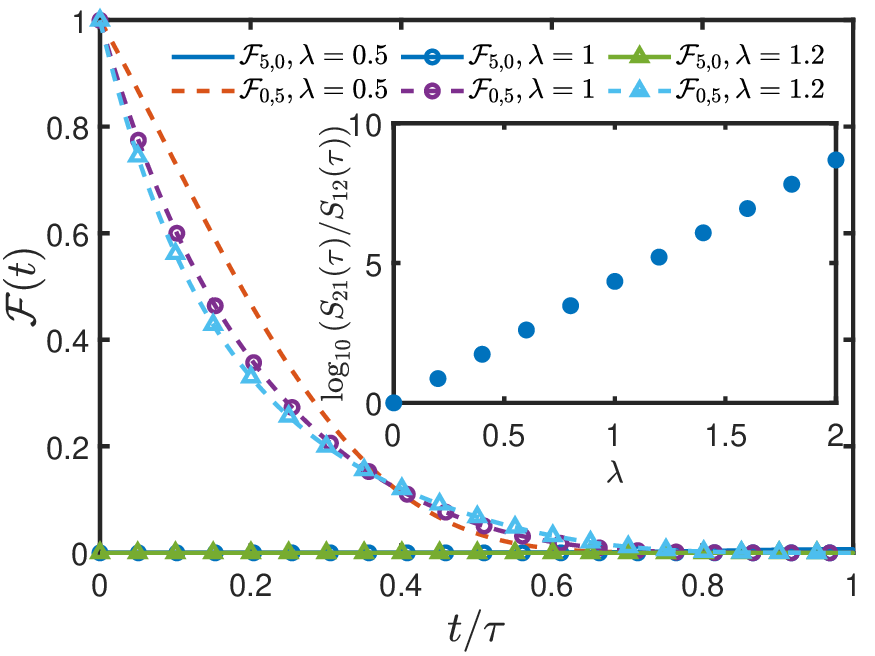}
\caption{Fidelity dynamics about the individual states $|5\rangle_a|0\rangle_b$ and $|0\rangle_a|5\rangle_b$ with various $\lambda$, the coupling strength or the dissipation rate of the system, under a $\mathcal{PT}$-symmetric-broken Hamiltonian. Inset: numerical results for the logarithm of the nonreciprocity $\log_{10}[S_{21}(\tau)/S_{12}(\tau)]$ as a function of $\lambda$. The other parameters are the same as Fig.~\ref{ConverFock}. The initial state of the system is $|\psi(0)\rangle=|0\rangle_a|5\rangle_b$.}\label{cohepa}
\end{figure}

In Fig.~\ref{cohepa}, we demonstrate the fidelity dynamics of the relevant states $|5\rangle_a|0\rangle_b$ and $|0\rangle_a|5\rangle_b$ for various $\lambda$. It is found that, although the state $|5\rangle_a|0\rangle$ is slightly populated, the population on the state $|0\rangle_a|5\rangle_b$ decreases monotonically with time, with a decay rate roughly proportional to $\lambda$ for $t\leq0.3\tau$. In particular, when $t=0.1\tau$, we have $\mathcal{F}_{0,5}=0.713$ for $\lambda=0.5$, $\mathcal{F}_{0,5}=0.601$ for $\lambda=1$, and $\mathcal{F}_{0,5}=0.557$ for $\lambda=1.2$. When $t=0.3\tau$, $\mathcal{F}_{0,5}=0.254$ for $\lambda=0.5$, $\mathcal{F}_{0,5}=0.212$ for $\lambda=1$, and $\mathcal{F}_{0,5}=0.199$ for $\lambda=1.2$. When $t\geq0.8\tau$, the population of the state $|0\rangle_a|5\rangle_b$ is almost vanishing for various $\lambda$. This result can be verified by the inset of Fig.~\ref{cohepa}, where the transmission coefficients demonstrate a dramatic nonreciprocal relation $S_{21}(\tau)\gg S_{12}(\tau)$ when $\lambda>0.5$, justifying the unidirectional perfect absorption in the cavity-magnonic system. Note that, for $\lambda=0$, $\log_{10}[S_{21}(\tau)/S_{12}(\tau)]=0$, i.e., $S_{21}(\tau)=S_{12}(\tau)$ implies the disappearance of the nonreciprocal behavior, reminiscent of the fact that, under the condition of Eq.~(\ref{condvanifi}) with $\lambda=0$, the non-Hermitian Hamiltonian in Eq.~(\ref{HamMCRe}) becomes Hermitian~\cite{Jin2026Bosonic}, i.e., $H(t)=iJ(t)a^\dagger b-iJ(t)b^\dagger a$. Our results then encompass the bidirectional perfect state transfer in the Hermitian bosonic system as a special case.

\section{Conclusion and discussion}\label{conclusion}

In summary, we propose a first-principles and versatile protocol for manipulating the general bosonic system through a non-Hermitian and time-dependent Hamiltonian. In sharp contrast to conventional methods, our theory describes the dynamics of non-Hermitian continuous-variable systems in the full Hilbert space through Heisenberg passages. It holds the quantum-jump terms in the Lindblad master equation and relies on the gauge freedom of differential manifolds (instantaneous frames) rather than the eigenspectrum. In the non-Hermitian cavity-magnonic system, our control protocol enables perfect and flexible transfer of arbitrary states between two modes and can be used to show unidirectional perfect absorption, irrespective of $\mathcal{PT}$ symmetry of the Hamiltonian or coefficient matrix and the exceptional points.

Practically, the system dynamics can be described by ancillary operators. By rotating from the time-dependent to the stationary ancillary representations, a purely geometric gauge potential emerges to shape the time evolution of the system, which is uniquely determined by the formation of ancillary operators. In the stationary representation, the assumption of the upper triangularized coefficient matrix of the non-Hermitian Hamiltonian provides a sufficient condition to decouple two Heisenberg passages for flexible control, e.g., perfect state transfer. Triangularization can be enforced by a sufficient number of tunable parameters and supported by Schur's decomposition theorem~\cite{Sheldon2024Linear}. Along the transitionless passages, arbitrary initial states can be perfectly transferred from the cavity mode to the magnon mode without artificial normalization, including but not limited to Fock states, coherent states, the superposition of Fock states, the cat states, and even thermal states. The resulting passages survive in both broken and unbroken phases of $\mathcal{PT}$ symmetry and with both avoiding and crossing EPs.

The present work is different from our previous UQC theory about the non-Hermitian discrete-variable systems~\cite{Jin2025NonHerm} in at least three aspects. (1) In Ref.~\cite{Jin2025NonHerm}, the triangularization assumption about the rotated Hamiltonian gives rise to a single nonadiabatic passage in either bra or ket space. The controllability of the present work depends on the triangularized coefficient matrix of Hamiltonian and it gives rise to two nonadiabatic Heisenserg passages in either bra or ket space. (2) The non-Hermitian Hamiltonian in Ref.~\cite{Jin2025NonHerm} is obtained by ignoring the quantum-jump process, but the present work retains the full information of the master equation. (3) The ansatz for the passages in discrete-variable systems are superposed of states, whereas those for the continuous-variable systems are superposed of operators.

When $H(t)=H^\dagger(t)$, the upper triangularization condition in Eq.~(\ref{UpTriaEq}) can become the commutation condition $[\mathcal{H}(t), \Pi^k]=0$ assumed by the UQC theory for continuous-variable systems under the Hermitian Hamiltonian~\cite{Jin2026Bosonic}. The latter was proved to be a {\it necessary and sufficient} condition to activate the ancillary operator $\mu_k(t)$ to be a useful Heisenberg passage in the closed $N$-mode bosonic system. If $k$ runs from $1$ to $N$, then $N$ passages can be constructed and the time-dependent Schr\"odinger equation is completely solved. However, the non-Hermitian Hamiltonian in the present work allows nonreciprocal state transfers between bosonic modes that cannot be attained in unitary evolution.

Essentially, this work much extends the application scope of our UQC framework and provides a powerful tool for controlling the open quantum continuous-variable systems.

\section*{Acknowledgments}

We acknowledge grant support from the National Natural Science Foundation of China (Grant No. U25A20199) and the ``Pioneer'' and ``Leading Goose'' R\&D of Zhejiang Province (Grant No. 2025C01028).

\appendix

\section{General recipe for ancillary operators}\label{recipe}

This Appendix provides a brief recipe for constructing the ancillary basis operators $\{\mu_k(t)\}$ for a general bosonic system of $N$ modes as well as the $N\times N$ unitary transformation matrix $\mathcal{M}^\dagger(t)$ in Eq.~(\ref{TimeAnci}). Rooted in the geometric structure of the manifold of $\mu_k(t)$, the adjoint matrix $\mathcal{M}^\dagger(t)$ admits the representation~\cite{Jin2025Entangling,Jin2026Bosonic}
\begin{equation}\label{unitary}
\mathcal{M}^\dagger(t)=\left[\vec{M}_1(t), \vec{M}_2(t), \ldots, \vec{M}_N(t)\right]^T
\end{equation}
with the $N$-dimensional row vectors
\begin{equation}\label{unitarVec}
\begin{aligned}
\vec{M}_1(t)&=\left(\cos\theta_1e^{i\frac{\alpha_1}{2}},-\sin\theta_1e^{-i\frac{\alpha_1}{2}},0,\ldots,0\right),\\
\vec{M}_k(t)&=\left[\cos\theta_ke^{i\frac{\alpha_k}{2}}\vec{b}_{k-1}(t),
-\sin\theta_ke^{-i\frac{\alpha_k}{2}},0,\ldots,0\right],\\ \cdots \\
\vec{M}_{N-1}(t)&=\Big[\cos\theta_{N-1}e^{i\frac{\alpha_{N-1}}{2}}\vec{b}_{N-2}(t), \\ &-\sin\theta_{N-1}e^{-i\frac{\alpha_{N-1}}{2}}\Big],\\
\vec{M}_N(t)&=\vec{b}_{N-1}(t),
\end{aligned}
\end{equation}
where $k$ runs from $2$ to $N-2$. Here $\vec{b}_k(t)$ is a $k+1$-dimensional bright vector
\begin{equation}\label{unitarVecBri}
\vec{b}_k(t)\equiv[\sin\theta_ke^{i\frac{\alpha_k}{2}}\vec{b}_{k-1}(t), \cos\theta_ke^{-i\frac{\alpha_k}{2}}],
\end{equation}
with $1\leq k\leq N-1$ and $\vec{b}_0(t)\equiv 1$. In Eqs.~(\ref{unitarVec}) and (\ref{unitarVecBri}), the time dependence of the parameters $\theta_k(t)$ and $\alpha_k(t)$ is treated implicitly for clarity. These parameters can be either time dependent or time independent.

With the definition in Eqs.~(\ref{TimeAnci}) and (\ref{unitary}), the rotation from the ancillary operators $\{\mu_k(t)\}$ to their stationary version $\mathcal{V}^\dagger(t)\mu_k(t)\mathcal{V}(t)\rightarrow\mu_k(0)$ can be performed by~\cite{Jin2026Bosonic}
\begin{equation}\label{unitaryV}
\mathcal{V}(t)=V_{\alpha_1}V_{\theta_1}V_{\alpha_2}V_{\theta_2}\cdots V_{\alpha_{N-1}}V_{\theta_{N-1}}=\prod_{k=1}^{N-1}V_{\alpha_k}V_{\theta_k},
\end{equation}
where
\begin{equation}\label{uniVat}
\begin{aligned}
V_{\alpha_k}(t)&=e^{-i\frac{\delta\alpha_k}{2}\left[b_{k-1}^\dagger(0)b_{k-1}(0)
-a_{k+1}^\dagger a_{k+1}\right]},\\
V_{\theta_k}(t)&=e^{-\delta\theta_k\left[e^{i\alpha_k(0)}a_{k+1}^\dagger b_{k-1}(0)-e^{-i\alpha_k(0)}b_{k-1}^\dagger(0)a_{k+1}\right]}
\end{aligned}
\end{equation}
with $\delta\alpha_k=\alpha_k(t)-\alpha_k(0)$ and $\delta\theta_k=\theta_k(t)-\theta_k(0)$.

\section{Derivation of activated passages from the upper-triangularized Hamiltonian}\label{SuffTrian}

This Appendix details the derivation from the upper triangular Hamiltonian $H_{\rm rot}(t)$ in Eq.~(\ref{HamrotDiaUp}) to the construction of the nonadiabatic passages in Eq.~(\ref{DycAncHN}).

In particular, the upper triangularized Hamiltonian $H_{\rm rot}(t)$ in Eq.~(\ref{HamrotDiaUp}) can be written in a matrix form as
\begin{widetext}
\begin{equation}\label{HamrotDiaUpMatr}
\begin{aligned}
H_{\rm rot}(t)&=\sum_{k=1}^N\sum_{m\geq k}^N\left[H^\mu_{km}(t)-\mathcal{A}_{km}(t)\right]\mu_k^\dagger(0)\mu_m(0)\equiv\sum_{k=1}^N\sum_{m\geq k}^N\mathcal{H}_{km}(t)\mu_k^\dagger(0)\mu_m(0)\\
&=\vec{\mu}_0^\dagger
\left(\begin{array}{ccccc}
	\mathcal{H}_{1,1}(t) & \mathcal{H}_{1,2}(t) & \cdots & \mathcal{H}_{1,N-1}(t) & \mathcal{H}_{1,N}(t) \\
	0 & \mathcal{H}_{2,2}(t) & \cdots & \mathcal{H}_{2,N-1}(t) & \mathcal{H}_{2,N}(t) \\
	\vdots & \vdots & \ddots & \vdots & \vdots \\
	0 & 0 & \cdots & \mathcal{H}_{N-1,N-1}(t) & \mathcal{H}_{N-1,N}(t) \\
	0 & 0 & \cdots & 0 & \mathcal{H}_{N,N}(t) \\
\end{array}\right)\vec{\mu}_0^T.
\end{aligned}
\end{equation}
\end{widetext}
The dynamics of an arbitrary operator $\tilde{\mathcal{O}}_S$ can be obtained by $\mathcal{O}_H(t)=V_{\rm rot}^\dagger(t)\tilde{\mathcal{O}}_SU_{\rm rot}(t)$ according to the non-Hermitian Heisenberg equation~\cite{Miao2016Investigation} with the evolution operators $U_{\rm rot}(t)$ and $V_{\rm rot}(t)$ given by Eq.~(\ref{Usch}). For the rotated Hamiltonian $H_{\rm rot}(t)$ in Eq.~(\ref{HamrotDiaUp}) or Eq.~(\ref{HamrotDiaUpMatr}), the dynamics of the ancillary operators $\mu_1^\dagger(0)$ and $\mu_N(0)$ can be written as
\begin{equation}\label{dynHeiApp}
\begin{aligned}
v_1^\dagger(t)&=V_{\rm rot}^\dagger(t)\mu_1^\dagger(0)U_{\rm rot}(t),\\
v_N(t)&=V_{\rm rot}^\dagger(t)\mu_N(0)U_{\rm rot}(t).
\end{aligned}
\end{equation}
By Eqs.~(\ref{HamrotDiaUpMatr}) and (\ref{dynHeiApp}), the time derivatives of $v_1^\dagger(t)$ and $v_N(t)$ can be decoupled as
\begin{equation}\label{dynHeiDer1App}
\begin{aligned}
&\frac{dv_1^\dagger(t)}{dt}=\frac{dV_{\rm rot}^\dagger(t)}{dt}\mu_1^\dagger(0)U_{\rm rot}(t)+V_{\rm rot}^\dagger(t)\mu_1^\dagger(0)\frac{dU_{\rm rot}(t)}{dt}\\
&=iV_{\rm rot}^\dagger(t)H_{\rm rot}(t)\mu_1^\dagger(0)U_{\rm rot}(t)-iV_{\rm rot}^\dagger(t)\mu_1^\dagger(0)H_{\rm rot}(t)\\
&\times U_{\rm rot}(t)=iV_{\rm rot}^\dagger(t)[H_{\rm rot}(t), \mu_1^\dagger(0)]U_{\rm rot}(t)\\
&=i\mathcal{H}_{1,1}(t)V_{\rm rot}^\dagger(t)\mu_1^\dagger(0)U_{\rm rot}(t)=i\mathcal{H}_{1,1}(t)v_1^\dagger(t),
\end{aligned}
\end{equation}
and
\begin{equation}\label{dynHeiDerNApp}
\begin{aligned}
&\frac{dv_N(t)}{dt}=\frac{dV_{\rm rot}^\dagger(t)}{dt}\mu_N(0)U_{\rm rot}(t)+V_{\rm rot}^\dagger(t)\mu_N(0)\frac{dU_{\rm rot}(t)}{dt}\\
&=iV_{\rm rot}^\dagger(t)H_{\rm rot}(t)\mu_N(0)U_{\rm rot}(t)-iV_{\rm rot}^\dagger(t)\mu_N(0)H_{\rm rot}(t)\\
&\times U_{\rm rot}(t)=iV_{\rm rot}^\dagger(t)[H_{\rm rot}(t), \mu_N(0)]U_{\rm rot}(t)\\
&=-i\mathcal{H}_{N,N}(t)V_{\rm rot}^\dagger(t)\mu_N(0)U_{\rm rot}(t)=-i\mathcal{H}_{N,N}(t)v_N(t),
\end{aligned}
\end{equation}
respectively. It is straightforward to find the other operators $\mu_{k}^\dagger(0)$ and $\mu_{k}(0)$, $k\neq1,N$, cannot be decoupled as $\mu_1^\dagger(0)$ and $\mu_N(0)$.

Equations~(\ref{dynHeiDer1App}) and (\ref{dynHeiDerNApp}) then yield the analytical solutions as
\begin{equation}\label{mu1SolvApp}
v_1^\dagger(t)=e^{if_1(t)}\mu_1^\dagger(0),  \quad v_N(t)=e^{-if_N(t)}\mu_N(0),
\end{equation}
respectively, with the global phases $f_k(t)\equiv\int_0^t\mathcal{H}_{k,k}(s)ds$, $k=1,N$. The dynamics of the operators in the picture governed by $H_{\rm rot}(t)$ is related to that in the original picture governed by $H(t)$ via the rotation~(\ref{HamTimeInAnci}). The dynamics of the ancillary operators $\mu_1^\dagger(0)$ and $\mu_N(0)$ in the original picture can then be obtained as
\begin{equation}\label{DycAncHApp}
\begin{aligned}
&\mu_1^\dagger(0)\rightarrow\mathcal{V}(t)v_1^\dagger(t)\mathcal{V}^\dagger(t)\\
=&e^{if_1(t)}\mathcal{V}(t)\mu_1^\dagger(0)\mathcal{V}^\dagger(t)=e^{if_1(t)}\mu_1^\dagger(t),
\end{aligned}
\end{equation}
and
\begin{equation}\label{DycAncHNApp}
\begin{aligned}
&\mu_N(0)\rightarrow\mathcal{V}(t)v_N(t)\mathcal{V}^\dagger(t)\\
=&e^{-if_N(t)}\mathcal{V}(t)\mu_N(0)\mathcal{V}^\dagger(t)=e^{-if_N(t)}\mu_N(t),
\end{aligned}
\end{equation}
respectively. They are exactly Eq.~(\ref{DycAncHN}) as the result of the {\it sufficient condition} proof started from the triangularization assumption. It is a {\it sufficient but not necessary} condition since inversely the decoupling of the interested passages does not necessarily yield the triangularization condition of the coefficient matrix.

\section{Derivation of non-Hermitian Hamiltonian}\label{HeffDerive}

This appendix provides a detailed derivation of the non-Hermitian Hamiltonian~(\ref{HamMC1}) from the Lindblad master equation. We emphasize that the quantum jump terms have been fully retained~\cite{Metelmann2015Nonreciprocal,Wang2019Nonreciprocity}, in contrast to those approaches only regarding the nonunitary evolution part after post-selection~\cite{Han2024Measuring}. In general, the dynamics of an open cavity magnonic system that is coupled to the traveling waves~\cite{Metelmann2015Nonreciprocal,Wang2019Nonreciprocity} can be described by
\begin{equation}\label{master}
\frac{d}{dt}\rho=-i[H_{\rm coh},\rho]+\eta\mathcal{L}[c]\rho+\beta\mathcal{L}[a]\rho+\chi\mathcal{L}[b]\rho,
\end{equation}
where $H_{\rm coh}=\omega_aa^\dagger a+\omega_bb^\dagger b+[J\exp(i\varphi)a^\dagger b+{\rm H.c.}]$ represents the eigen-energies of the two modes and the coherent coupling between them. The Lindblad superoperators are defined as $\mathcal{L}[o]\rho=o\rho o^\dagger-\{o^\dagger o,\rho\}/2$, $o=c,a,b$. The first superoperator is about $c\equiv ua+\exp{(i\Theta)}vb$ with weight factors $u$ and $v$. It implies the cooperative interactions between the two modes and the traveling waves with a damping rate $\eta$. $\Theta=0$ or $\Theta=\pi$ is determined by the propagation direction of the traveling waves. The second and the third superoperators are associated with the individual dissipations of the modes $a$ and $b$, with the damping rates $\beta$ and $\chi$, respectively.

Under Eq.~(\ref{master}), the Schr\"odinger-picture operator $\mathcal{O}_S$ is connected to the Heisenberg-picture operator $\mathcal{O}_H(t)$ as
\begin{equation}\label{masterSH}
\begin{aligned}
&{\rm Tr}\left[\mathcal{O}_S\dot{\rho}(t)\right]={\rm Tr}\Big[\mathcal{O}_S\Big(-i[H_{\rm coh},\rho]+\eta\mathcal{L}[c]\rho+\beta\mathcal{L}[a]\rho\\
&+\chi\mathcal{L}[b]\rho\Big)\Big]={\rm Tr}\Big[\Big(i[H_{\rm coh},\mathcal{O}_S]+\eta\mathcal{L}^\dagger[c]\mathcal{O}_S+\beta\mathcal{L}^\dagger[a]\mathcal{O}_S\\
&+\chi\mathcal{L}^\dagger[b]\mathcal{O}_S\Big)\rho\Big]={\rm Tr}\left[\dot{\mathcal{O}}_H(t)\rho(0)\right],
\end{aligned}
\end{equation}
where the Hermitian conjugate superoperator $\mathcal{L}^\dagger[o]\mathcal{O}_S\equiv o^\dagger\mathcal{O}_So-\{o^\dagger o,\mathcal{O}_S\}/2$. The derivation from the second line to the third one in Eq.~(\ref{masterSH}) has used the cyclic property of the trace. Thus, the time derivative of $\mathcal{O}_H(t)$ can be expressed by the adjoint Lindblad master equation as
\begin{equation}\label{OHdynamic}
\frac{d}{dt}\mathcal{O}_H(t)=i[H_{\rm coh},\mathcal{O}_S]+\eta\mathcal{L}^\dagger[c]\mathcal{O}_S+\beta\mathcal{L}^\dagger[a]\mathcal{O}_S
+\chi\mathcal{L}^\dagger[b]\mathcal{O}_S.
\end{equation}

Using Eq.~(\ref{OHdynamic}), the dynamics of the modes $a$ and $b$ can be obtained as
\begin{equation}\label{abdynamic}
\begin{aligned}
\frac{d}{dt}a(t)&=-i\omega_aa-\gamma_aa-\left(iJ+e^{i\Theta}\Gamma\right)b,\\
\frac{d}{dt}b(t)&=-i\omega_bb-\gamma_bb-\left(iJ+e^{i\Theta}\Gamma\right)a,
\end{aligned}
\end{equation}
under the conditions of $\Gamma=\eta uv$, $\gamma_a=\eta v^2+\beta$, and $\gamma_b=\eta u^2+\chi$. The system dynamics described by Eq.~(\ref{abdynamic}) is equivalent to that governed by the Hamiltonian~(\ref{HamMC1}) with $\varphi_a=0$.

\section{Illustrative example for the upper triangularization condition}\label{Exam}

This Appendix provides an example about the upper-triangular condition in Eq.~(\ref{UpTriaEq}) by using an open two-mode bosonic system governed by the non-Hermitian Hamiltonian~(\ref{HamMC}).

In particular, the Hamiltonian~(\ref{HamMC}) can be written in terms of the coefficient matrix $H^a(t)$ as
\begin{equation}\label{TriApp}
\begin{aligned}
H(t)&=\begin{pmatrix}a^\dagger&b^\dagger\end{pmatrix}H^a(t)\begin{pmatrix}a\\ b\end{pmatrix}\\
&=\begin{pmatrix}a^\dagger&b^\dagger\end{pmatrix}\begin{pmatrix}\frac{\Delta}{2}
-i\gamma_ae^{i\varphi_a}&J(t)e^{i\varphi}+i\Gamma\\ J(t)e^{-i\varphi}+i\Gamma & -\frac{\Delta}{2}-i\gamma_b\end{pmatrix} \begin{pmatrix}a\\ b\end{pmatrix}.
\end{aligned}
\end{equation}
Using the unitary transformation matrix in Eq.~(\ref{unitary}) with $N=2$, Eq.~(\ref{TriApp}) can be expressed in the time-dependent ancillary representation as
\begin{equation}\label{TriAnciApp}
H(t)=\begin{pmatrix}\mu_1^\dagger(t)&\mu_2^\dagger(t)\end{pmatrix}H^\mu(t)\begin{pmatrix}\mu_1(t)\\ \mu_2(t)\end{pmatrix},
\end{equation}
where the rotated coefficient matrix $H^\mu(t)$ can be obtained as
\begin{equation}\label{HrotMatApp}
\begin{aligned}
H^\mu(t)=\mathcal{M}^\dagger(t)H^a(t)\mathcal{M}(t)=\begin{pmatrix}H^\mu_{11}(t)&H^\mu_{12}(t)\\ H^\mu_{21}(t) & H^\mu_{22}(t)\end{pmatrix}
\end{aligned}
\end{equation}
with the elements
\begin{equation}\label{MrotMatCoe}
\begin{aligned}
H_{11}^\mu(t)&=\left[\frac{\Delta}{2}-i\gamma_ae^{i\varphi_a}\right]\cos^2\theta(t)-\left(\frac{\Delta}{2}+i\gamma_b\right)\sin^2\theta(t)\\
&-\left[J(t)\cos(\varphi+\alpha)+i\Gamma\cos\alpha\right]\sin2\theta(t),\\
H_{22}^\mu(t)&=\left[\frac{\Delta}{2}-i\gamma_ae^{i\varphi_a}\right]\sin^2\theta(t)-\left(\frac{\Delta}{2}+i\gamma_b\right)\cos^2\theta(t)\\
&+\left[J(t)\cos(\varphi+\alpha)+i\Gamma\cos\alpha\right]\sin2\theta(t),\\
H_{12}^\mu(t)&=\left[\Delta-i\gamma_ae^{i\varphi_a}+i\gamma_b\right]\sin\theta(t)\cos\theta(t)\\
&+\Big[J(t)\cos(\varphi+\alpha)+i\Gamma\cos\alpha\Big]\cos2\theta(t)\\
&+i\left[J(t)\sin(\varphi+\alpha)+i\Gamma\sin\alpha\right],\\
H_{21}^\mu(t)&=\left[\Delta-i\gamma_ae^{i\varphi_a}+i\gamma_b\right]\sin\theta(t)\cos\theta(t)\\
&+\Big[J(t)\cos(\varphi+\alpha)+i\Gamma\cos\alpha\Big]\cos2\theta(t)\\
&-i\left[J(t)\sin(\varphi+\alpha)+i\Gamma\sin\alpha\right].
\end{aligned}
\end{equation}
In the stationary representation with respect to $\mathcal{V}(t)$ in Eq.~(\ref{unitaryV}) with $N=2$, the Hamiltonian can be transformed as
\begin{equation}\label{TriTrans}
\begin{aligned}
H_{\rm rot}(t)&=\mathcal{V}^\dagger(t)H(t)\mathcal{V}(t)-i\mathcal{V}^\dagger(t)\frac{d\mathcal{V}(t)}{dt}\\
=&\begin{pmatrix}\mu_1^\dagger(0)&\mu_2^\dagger(0)\end{pmatrix}\left[H^\mu(t)-\mathcal{A}(t)\right]\begin{pmatrix}\mu_1(0)\\ \mu_2(0)\end{pmatrix},
\end{aligned}
\end{equation}
with the gauge potential
\begin{equation}\label{Amatrix}
\mathcal{A}(t)=\frac{1}{2}\begin{pmatrix}\dot{\alpha}(t)\cos2\theta(t)&\dot{\alpha}(t)\sin2\theta(t)+i2\dot{\theta}(t)\\ \dot{\alpha}(t)\sin2\theta(t)-i2\dot{\theta}(t)& -\dot{\alpha}(t)\cos2\theta(t)\end{pmatrix}.
\end{equation}
Using Eqs.~(\ref{MrotMatCoe}), (\ref{TriTrans}), and (\ref{Amatrix}), the upper triangularization of $H_{\rm rot}(t)$, i.e., $\mathcal{H}_{21}=0$, yields the constraint conditions for $J(t)$ and $\Delta(t)$ in Eq.~(\ref{Hconstr}).

The triangularization condition~(\ref{UpTriaEq}) for general coefficient matrices of arbitrary dimensions, which lacks a purely theoretical or mathematical justification, should be regarded as a physical choice that defines the constraints. To a certain degree, it is mathematically supported by the Schur's decomposition theorem~\cite{Sheldon2024Linear}, which states that every operator on a finite-dimensional complex inner product space has an upper-triangular matrix with respect to some orthonormal basis.

\bibliographystyle{apsrevlong}
\bibliography{ref}

\end{document}